\documentclass{aastex62}

\usepackage{amsmath}
\usepackage{stackengine}

\newcommand{\noop}[1]{}

\newcommand{\Myr} {\mbox{$~\text{Myr}$}}
\newcommand{\AU} {\mbox{$~\text{AU}$}}
\newcommand{\pc} {\mbox{$~\text{pc}$}}

\newcommand{\MSun} {\mbox{$M_{\odot}$}}
\newcommand{\MJup} {\mbox{$~M_{Jup}$}}
\newcommand{\Mcluster} {\mbox{$~M_{\text{cl}}$}}
\newcommand{\Mstars} {\mbox{$~M_{\text{stars}}$}}
\newcommand{\Mgas} {\mbox{$~M_{\text{gas}}$}}
\newcommand{\rhoM} {\mbox{$~M_{\odot}~\text{pc}^{-3}$}}

\begin{document} 

   \title{How do disks and planetary systems in {high-mass} open clusters differ from those around field stars?}

%\received{January 1, 2018}
%\revised{January 7, 2018}
%\accepted{\today}

 \correspondingauthor{Susanne Pfalzner}
\email{spfalzner@mpifr.de}
   
   \author{Kirsten Vincke}
   \affiliation{Max Planck Institute for Radio Astronomy, Auf dem H\"ugel 69, 53121 Bonn, Germany}

  \author{Susanne Pfalzner}
  \affiliation{Max Planck Institute for Radio Astronomy, Auf dem H\"ugel 69, 53121 Bonn, Germany}

 \begin{abstract}

Only star clusters that are sufficiently compact and massive survive largely unharmed beyond $10\Myr$. However, their compactness means a high stellar density which can lead to strong gravitational interactions between the stars. As young stars are often initially surrounded by protoplanetary disks and later on potentially by planetary systems, the question arises to what degree these strong gravitational interactions influence planet formation and the properties of planetary systems.   Here, we perform simulations of the evolution of compact high-mass clusters like Trumpler 14 and Westerlund 2 from the embedded to the gas-free phase and study the influence of stellar interactions. We concentrate on the development of the mean disk size in these environments.  Our simulations show that in high-mass open clusters $80-90\%$ of all disks/planetary systems should be smaller than $50\AU$ just due to the strong stellar interactions in these environments. Already in the initial phases, 3-4 close fly-bys lead to typical disk sizes within the range of $18-27\AU$.   Afterwards, the disk sizes are altered only to a small extent. Our findings agree with the recent observation that the disk sizes in the 
once dense environment of the Upper Scorpio OB association,  NGC~2362,  and h/$\chi$Persei are at least three times smaller in size than, for example, in Taurus. We conclude that the observed planetary systems in high-mass open clusters should also be on average smaller than those found around field stars; in particular, planets on wide orbits are expected to be extremely rare in such environments.

 \end{abstract}


\section{Introduction}
\label{sec:intro}

  Unlike most clusters/associations in the solar neighborhood, which often dissolve within 10 Myr \citep{Porras_et_al_2003,Lada_Lada_2003}, some clusters can remain intact for hundreds of Myr and more. 
 These clusters are characterized to be compact (1-3 pc) and relatively massive - properties they have inherited from their formation phase when they were likely even more compact ($0.1-0.5\pc$). Clusters like NGC 3603, Arches, Trumpler 14 and Westerlund 2 ($\sim 2\Myr$) are thought to be younger counterparts.\footnote{Currently the nomenclature is ambiguous as these two groups are sometimes referred to as compact vs. extended/loose/leaky clusters or clusters vs. associations. In the following, we will use the terms compact and extended clusters.}

Given their small sizes and large masses ($M_c > 10^4$ \MSun) \citep{Figer_2008} the stellar density in such clusters is very high; for example, up to $\sim 2*10^{5}\rhoM$ \citep{Espinoza_Selman_Melnick_2009} in the central areas of Arches. Initially, the stars are mostly surrounded by disks from which planetary systems may form. 
  However, the high stellar density means that protoplanetary disks in such dense environments can be influenced by external processes like external photoevaporation \citep{Johnstone_Hollenbach_Bally_1998, Stoerzer_Hollenbach_1999, Scally_Clarke_2002, Matsuyama_Johnstone_Hartmann_2003, Johnstone_et_al_2004, Adams_et_al_2006, Alexander_Clarke_Pringle_2006, Ercolano_et_al_2008, Gorti_Hollenbach_2009, Drake_et_al_2009, Clarke_Owen_2015, Haworth_Clarke_Owen_2015} or gravitational interactions \citep[e.g.][]{Clarke_Pringle_1993, Hall_1997, Scally_Clarke_2001}. The latter can also alter already formed planetary systems \citep[see, for example,][]{fuente:97,laughlin:98,spurzem:09,shara:16}. In extreme cases these two processes can lead to disk destruction, decreasing the disk lifetime in such clusters. However, more frequrntly the disk is truncated leading to a smaller disk size.
Here, we concentrate on the effect of fly-bys on the disk size in compact clusters that develop into long-lived clusters, because this effect is present throughout all the cluster stages and it affects protoplanetary disks as well as already formed planetary systems. Taking Trumpler 14 and NGC 3603 as templates means that we look at the high mass end of clusters ($M_c \geq 10^4$ \MSun), which are sometimes referred to as starburst clusters. Many open clusters have somewhat lower masses. A comparison to lower mass compact clusters will be given in section 4.3.

    \begin{figure}[t!]
      \centering
      \includegraphics[width=0.45\textwidth]{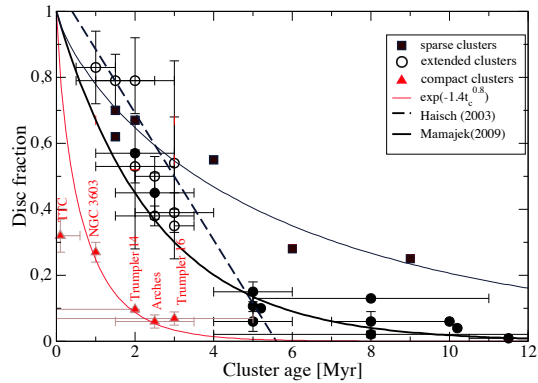}
      \caption{Disk fraction as a function of cluster age for sparse, extended and compact clusters. Values from \citet{pfalzner:14} and \citet{Stolte_et_al_2015}.}
      \label{fig:cluster_density}
    \end{figure}

  % Table 2.
\setcounter{table}{1}
  \begin{table}h!]
\renewcommand{\thetable}{\arabic{table}}
 %   \centering
    \caption{Observed properties of planets in open clusters.}\label{tab:observed_planetary_systems_in_open_clusters}
    \begin{tabular}[t]{llllllll} 
\hline
\hline
      \multicolumn{3}{l}{Cluster properties}                                                                      & \multicolumn{4}{l}{Planet properties}                                                         & References        \\ 
                                   Name        & $t_{\text{cl}}$    & $M_{\text{cl}} $    & Host star      & $m_{\text{pl}}$        & $a_{\text{pl}}$         & $e_{\text{pl}}$           &                   \\
		  & [Myr]              & [$M_{\odot}$]    &                & [$\MJup$]              & [AU]                    &                           &                   \\ 
%\vspace{0.3em}\\
\hline
\decimals
      %------------------------------------------------------------------------------------------------------------------------------------------------------------------------------
      NGC~4349   & $200$             & ---                & No.~127        & $19.8^{a)}$            & $2.38$                  & $0.19$                    & (1)               \\ 
      %------------------------------------------------------------------------------------------------------------------------------------------------------------------------------
      NGC~2632   & $578 \pm 12$      & $550 \pm 40$       & Pr 0201        & $0.540 \pm 0.039^{a)}$ & $0.057$                 & $0$                       & (2), (3), (4)     \\
      (Praesepe) &                   &                    & Pr 0211        & $1.8 \pm 0.1^{a), b)}$ & $0.03 \pm 0.01^{b)}$    & $0.011 \pm 0.011^{b)}$    & (4), (5)          \\ \vspace{0.3em}
		 &                   &                    & Pr 0211        & $7.79 \pm 0.33^{a)}$   & $5.5^{+3.0, c)}_{-1.4}$ & $0.71 \pm 0.11$           & (5)               \\ 
	 &                   &                    & K2-95        &                                        & 0.069 & 0.16      & (20)               \\ 
	 &                   &                    & K2-100       &                                        & 0.029 & 0.24      & (20)               \\ 
	 &                   &                    & K2-101       &                                        & 0.11 & 0.10     & (20)               \\ 
	 &                   &                    & K2-102        &                                        & 0.083& 0.10     & (20)               \\ 
	 &                   &                    & K2-103     &                                        & 0.013 & 0.18    & (20)               \\ 
	 &                   &                    & K2-104       &                                        & 0.025 & 0.18     & (20)               \\ 
      %------------------------------------------------------------------------------------------------------------------------------------------------------------------------------
      Hyades     & $625 \pm 50$      & $300-400$          & $\epsilon$~Tau & $7.6 \pm 0.2$          & $1.93 \pm 0.03$         & $0.151 \pm 0.023$         & (6), (7)          \\ \vspace{0.3em}
		       &                   &                    & HD~285507      & $0.917 \pm 0.033^{a)}$ & $0.06^{d)}$             & $0.086^{+0.018}_{-0.019}$ & (8)               \\ \vspace{0.3em}
		       &                   &                    & K2-136      &                                       & Period: 7.97d           & $<$ 0.72& (17), (18), (19)               \\
                       &                   &                    &  K2-136                &                                       &  Period:17.31d       & $<0.47$ &  (17), (18), (19)              \\ 
                      &                   &                    &   K2.136                &                                       &  Period: 25.57d        & $<0.75$ &  (17), (18), (19)              \\ \vspace{0.3em}
      %------------------------------------------------------------------------------------------------------------------------------------------------------------------------------
      NGC~2423   & $750$             & ---                & No.~3          & $10.6^{a)}$            & $2.10$                  & $0.21$                    & (1)               \\ 
      %------------------------------------------------------------------------------------------------------------------------------------------------------------------------------
      NGC~6811   & $1\,000 \pm 170$  & $\sim 3\,000^{e)}$ & Kepler~66      & $\le 0.06$             & $0.1352 \pm 0.0017$     & ---                       & (9), (10)         \\ \vspace{0.3em}
		 &                   &                    & Kepler~67      & $\le 0.06$             & $0.1171 \pm 0.0015$     & ---                       & (10)              \\ 
      Rup 147      & 3,000                     &                           & K2-231          & $7^{+5}_{-3}$               &   & &  (16)\\
      %------------------------------------------------------------------------------------------------------------------------------------------------------------------------------
      NGC~2682   & $3\,500 - 4\,000$ & $1\,080$           & YBP1194        & $0.34 \pm 0.05^{a)}$   & $0.07$                  & $0.24 \pm 0.08$           & (11), (12), (13)  \\
      (M67)           &                               &                           & YBP1514        & $0.40 \pm 0.11^{a)}$   & $0.06^{d)}$          & $0.39 \pm 0.17$           & (13)              \\
		          &                               &                           & SAND364       & $1.54 \pm 0.24^{a)}$   & $0.53^{d)}$          & $0.35 \pm 0.08$           & (13)              \\ 
		          &                               &                           & YBP401          & $0.46 \pm 0.05^{a)}$   & $0.05^{d)}$          $0.15 \pm 0.08$              & (14), (15)      \\ 
      %-------------------------------------------------------------------------------------------------------------------------------------------------------------------------
\hline
\end{tabular}
\tablecomments{Column~1 indicates the cluster name, $t_{\text{cl}}$ its age and $M_\text{cl}$ its mass, Col.~4 gives the name of the planet-hosting star,
$m_{\text{pl}}$ the planet mass, $a_{\text{pl}}$ its semi-major axis, $e_{\text{pl}}$ its eccentricity, and Col.~8 the references below.
   References: (1) \cite{Lovis_Mayor_2007}, (2) \cite{Delorme_et_al_2011}, (3) \cite{Kraus_Hillenbrand_2007}, (4) \cite{Quinn_et_al_2012}, (5)
\cite{Malavolta_et_al_2016}, (6) \cite{Perryman_et_al_1998}, (7) \cite{Sato_et_al_2007}, (8) \cite{Quinn_et_al_2014}, (9) \cite{Janes_et_al_2013}, (10) \cite{Meibom_et_al_2013}, (11) \cite{Sarajedini_Dotter_Kirkpatrick_2009}, (12) \cite{Richer_et_al_1998}, (13) \cite{Brucalassi_et_al_2014}, (14)
\cite{Pietrinferni_et_al_2004}, (15) \cite{bruca:16}, (16) \cite{curtis:18}, (17) \cite{livingston:18}, (18) \cite{livingston:18}, (19) \cite{ciardi:18}, (20) \cite{mann:16}.
Comments: a) given as $m_{\text{pl}}*\sin(i)$, where $i$ is the inclination between the orbital plane
and the line of view; b) combined planet properties from (4) and (5), errors give differences between both data sets; c) note the large error in fitted period:
$P=4\,850^{+4560}_{-1750}$ days; d) calculated from given orbital periods; e) crude estimate for $6\,000$ cluster members as suggested by (10).
}
  \end{table}

There are only a few surveys providing disk sizes in compact young massive clusters, because determining the disk sizes in these environments is challenging.  One reason is that most of these high-mass compact clusters are located at relatively large distances, for example, Trumpler 14 is  2.7 kpc away. In addition, the compactness and high mass of these clusters means that crowding is a major issue. A further problem is that due to their high mass these clusters contain many massive stars which dominate the radiation. However, disk frequencies are better constrained by observations than the disk sizes.  For example, in the young compact clusters Arches ($2.5 \pm 0.5\Myr$) and the Quintuplet ($4 \pm 1\Myr$) disk frequencies of $6 \pm 2\%$ and $4.0 \pm 0.7\%$, respectively, have been observed \citep{Stolte_et_al_2010, Stolte_et_al_2015}. These are considerably smaller than the ones found in less dense environments such as the Orion nebula cluster with approximately 70\% \citep[][]{Hillenbrand_et_al_1998}, as shown in Fig.1. Therefore, it can be concluded that a dense environment has a strong effect on the disks. Whether the lower disk fraction in compact clusters is mainly due to the effect of fly-bys or external photo-evaporation is an open question, as will be discussed in section 4.3. Here, we concentrate on the effect of fly-bys on the disk size, but future investigations should also consider the effects of external photoevaporation which have been neglected here.
 
 As protoplanetary disks are the pre-requisite for planet formation, there have been speculations whether planets could at all form in such harsh environments.
Early surveys of exoplanets in {long-lived open and globular} clusters could only give upper limits for the portion of stars having a Jupiter-like companion \citep[see e.g.][]{Paulson_Cochran_Hatzes_2004, Mochejska_et_al_2006, Pepper_et_al_2008, Hartman_et_al_2009, van_Saders_Gaudi_2011}.
 So far about 20  planets around main-sequence stars have been found in seven different open clusters, see Table~\ref{tab:observed_planetary_systems_in_open_clusters}, among them even planetary systems containing at least two planets. In addition, some planetary candidates have been found \citep{bruc:17}. Apart from the planets listed in Table 1, there is also indirect evidence for the presence of so-far undetected planets through 
external metal pollution of white dwarfs; for example, in the Hyades cluster \citep{farihi:13,zuckerman:13}. However, most of these clusters have a lower mass than the ones studied here. In addition, let us consider the system PSR B1620–26 which consists of a millisecond pulsar-white dwarf binary surrounded by a Jupiter-mass planet in a 40 yr orbit \citep{Thorsett_93} in the globular cluster NGC 6121 (M4). Its formation has been related to dynamical interactions that occurred in the cluster \citep{Sigurdsson_03}. In summary, at least some planetary systems are able to survive in such dense stellar environments for many Gyr. 
Given the low number of known planets in clusters it is still unclear whether planets are as frequent in clusters as in the field \citep{van_Saders_Gaudi_2011, Meibom_et_al_2013}.

As mentioned above, we concentrate here on  the typical disk and planetary-system (DPS) size in dense clusters as a result of fly-bys.  More specifically,
 we address the question whether the sizes of the DPS in open clusters differs from those found around field stars, and if so, how to quantify these differences in size.\footnote{DPS sizes refer here and in the following to the disk radius rather than the DPS diameter.} 

We tackle these questions by performing numerical simulations of young compact clusters and following their evolution over the first $10\Myr$ of their existence. Afterwards, we have analyzed  the data to study the influence of the cluster environment on protoplanetary disks.  There have been numerous studies investigating the effect of the gravitational interactions on protoplanetary disks and planetary systems in cluster environments. However, many of them concentrated on the fly-by  rate or disk frequency \citep{Olczak_Pfalzner_Spurzem_2006, Pfalzner_Olczak_Eckart_2006, Olczak_Pfalzner_Eckart_2010, Craig_Krumholz_2013, Steinhausen_Pfalzner_2014}. In addition,  they mostly focus on the less dense clusters typical for the solar neighborhood \citep{Rosotti_et_al_2014, Vincke_Breslau_Pfalzner_2015, Portegies_16,Vincke_Pfalzner_2016}. 
A recent study investigates the environmental effects during the interesting star formation phase  \citep{bates:18}, however, here again a less dense environment is considered and only the first 0.5 Myr have been modeled. Studies that investigate the effect on disk sizes or planetary system sizes including dense clusters were recently carried out by  \citet{Portegies_Zwart_2015} and \citet{winter:18a}. However, \citet{Portegies_Zwart_2015} only gives a rough analytical estimate of an unaffected zone and \citet{winter:18a} look at the current density in the considered clusters, but do not take into account the temporal development of the clusters. 

However, dense clusters evolve along well defined radius-age and mass-age tracks, changing their stellar density by orders of magnitude during the first 10 Myr \citep[][]{Pfalzner_2009,Pfalzner_Kaczmarek_2013b}. In stark contrast to their less dense counterparts like the Taurus or even the ONC, in dense clusters only a few stars become unbound due to gas expulsion at the end of the star formation process. Actually, stellar encounters are the main driving force of cluster expansion \citep{Pfalzner_Kaczmarek_2013b}, so stellar density  in these clusters steadily declines with age. The effect of cluster expansion on disk sizes has so far not been modeled for dense clusters. Only the effect of a more or less fixed density environment was modeled in above mentioned studies. However, expansion naturally occurs during the development of such clusters which might influence crucially the effect on DPS.

  Ideally one would like to model the cluster dynamics and resolve at the same time the evolution of the disk that initially surrounds  each star. There have been first attempts in such a direct method \citep{Rosotti_et_al_2014} at a fixed density.  However, these simulations are limited to a low number (100) of stars, that have equal mass  and are in a lower density environment. Therefore much fewer interactions happen and only the first $\sim 0.5\Myr$ can be modeled due to computational limitations. By contrast, modelling the compact cluster progenitors requires to treat at least 1000 stars with an approximately 1000 times higher stellar density, and it is essential that the stellar masses are chosen according to the IMF, as the effect of gravitational focusing is very important in this context \citep{Pfalzner_Olczak_Eckart_2006}.   
 Therefore, we perform here a two step approach, where we first model the cluster dynamics while recording all the fly-bys, and then post-process the data to determine the effect of the close fly-bys on the DPS.

 In addition, the model adopted here has as main advantage that it can be used equally for the disk as well as the planetary system stage, referred to here as DPS. Thus we do not need to know how fast planet formation actually happens, which is still a major point of discussion. For disks the simulation particles are representative for the mass distribution of the dust, whereas for planetary systems they represent the parameter space where planets potentially move on circular orbits before the fly-by.
  
 An additional difference to previous studies is that most of them considered all encounters to be prograde and coplanar.  These type of fly-bys are the most destructive ones and the determined losses can be interpreted as the upper limit of the effect of fly-bys on DPS in general \citep{Clarke_Pringle_1993,Heller_1995,Hall_1997,Bhandare_Breslau_Pfalzner_2016}. Here we will investigate the more realistic situation of randomly orientated fly-bys. In summary, the study presented here differs from previous work as it (i) models all phases of the cluster development (embedded, expansion and semi-equilibrium), (ii) treats dense clusters that will like develop into long-lived open clusters and (iii) includes fly-bys of arbitrary orientation.

  In Sect.~\ref{sec:method} we will describe the cluster simulations and the disk-size determination in detail. 
  Afterwards, we present our results on the effect of the dense cluster environment on the disk-size distributions in Sect.~\ref{sec:results}.  The
  assumptions that we have made in our set-up will be discussed in Sect.~\ref{sec:discussion}. In Sect.~\ref{sec:summary_and_conclusion}, we will discuss the differences that can be expected when comparing planetary systems in open clusters to those around field stars, and give a short summary.

\section{Method}
\label{sec:method}

  As mentioned in the introduction, we use a two-step approach to determine the effect of fly-bys on the DPS around stars in dense clusters. First, while simulating the cluster dynamics we simultaneously record the fly-by history.
  This is similar to what has been done in our earlier work described in \cite{Vincke_Pfalzner_2016} and \cite{Pfalzner:18a} where more details of the method can be found.

  %---------------------------------------------------------------------------------------------------------------------------------------------------------------------------------

  \subsection{Cluster dynamics}
  \label{sec:cluster_dynamics}
  
    The cluster simulations are performed using the code Nbody6++GPU \citep{Aarseth_1973, Spurzem_1999, Aarseth_2003, Wang}, which is an optimized version of NBODY6++ with hybrid parallelization methods (MPI, GPU, OpenMP, and AVX/SSE) to accelerate large direct N-body simulations.
 In contrast to \cite{Vincke_Pfalzner_2016},  we focus here on very dense - potentially long-lived - clusters representative for the case of  Arches or Westerlund 2. As such they are more massive than the special case of M44 studied in \cite{Pfalzner:18a}. We start at that point in time when star formation is completed. This means that the times given are not necessarily equivalent to the cluster age as the star formation phase is not covered in our simulations.
    We study two cases: one without an embedded phase (C0) and the other one where we have considered a $1\Myr$ long embedded phase (C1). For the other properties of the simulated clusters, see Table~\ref{tab:set-up}. 

In order to study to what extent the planetary systems in compact/open clusters resemble or differ from those found around field stars and those formed in less dense clusters, we compare our results to our earlier work, where we modeled less dense clusters typical for the solar neighborhood. The parameters of this cluster (model E52) are also given in Table~\ref{tab:set-up}. Most importantly, E52 has the same initial mass as C0 and C1, but the half-mass radius is larger ($1.3\pc$) and the star formation efficiency (SFE), that is the fraction of gas in the cluster which is turned into stars,  smaller ($30\%$ compared to 70\%). In this case the embedded phase, $t_{emb}$, was assumed to last $2\Myr$.

  %---------------------------------------------------------------------------------------------------------------------------------------------------------------------------------

    \subsubsection{Cluster initial conditions}
    \label{sec:cluster_initial_conditions}
 
The starting point of our simulations, $t_0 = 0$, corresponds to a fully formed cluster. In NBody6++ the gas is not modeled explicitly but just as a background potential.
The clusters are set up with an initial half-mass radius of $0.2\pc$ that is   
%roughly to that of the Arches cluster \citep{Stolte_et_al_2010}. Such a size might not only be applicable to the Arches, but 
typical for compact clusters at the start of the expansion phase \citep{Pfalzner_Kaczmarek_2013b}.
      The SFE is assumed to be $70\%$. There are two reasons why it is necessary to assume such a high SFE value, i) observations  hint at much higher SFEs for compact clusters \citep{henault,rochau:10,Cottaar_12} than for those in the solar neighborhood  and ii) if one interprets the size-age relation in compact clusters as temporal sequence because  this demands also such a high SFE \citep[][]{Pfalzner_Kaczmarek_2013b}.
    
      Usually, the stellar density distribution is modeled either as a King or Plummer profile \citep[cf.][]{Bastian_Goodwin_2006, Rosotti_et_al_2014, Banerjee_Kroupa_2015, Wang}.
      Here we choose for a modified King profile for the stars and a corresponding Plummer profile for the gas which reflects the situation in observed clusters \citep{Espinoza_Selman_Melnick_2009, Steinhausen_phd_2013}.\footnote{Note that the simulations are not supposed to exactly reproduce the Arches cluster, but to represent compact clusters in general.} The total cluster mass is given by \mbox{\Mcluster=\Mstars+\Mgas} with $M_{\text{stars}}$ being the stellar component of the cluster and $M_{\text{gas}}$ the gas mass. It follows:
      \begin{equation}
	M_{\text{gas}} = \frac{M_{\text{stars}} (1-\text{SFE})}{\text{SFE}}.
	\label{eq:gas_mass}
      \end{equation}
      The stellar masses were sampled from an IMF \citep{Kroupa_2002} with a lower limit of $0.08\MSun$ and an upper limit of $150\MSun$ and the velocities were sampled from a Maxwellian distribution.
    
  %- - - - - - - - - - - - - - - - - - - - - - - - - - - - - - - - - - - - - - - - - - - - - - - - - - - - - - - - - - - - - - - - - - - - - - - - - - - - - - - - - - - - - - - - -  

    \subsubsection{Fly-by frequency}
    \label{sec:fly-by_frequency}
 
      In all cases we assume the clusters to be initially in virial equilibrium.  The gas expulsion is modeled as being instantaneous, as $t_{\text{gas}} < t_{\text{emb}} < 1\Myr$. 
      It is not clear how long the embedded phase of compact clusters lasts. 
      None of the compact clusters younger than $3\Myr$, like for example the Arches cluster, show signs of considerable amounts of gas \citep[see][]{Pfalzner_2009}. Thus, to date, no unambiguous embedded precursor cluster has been identified.  From the absence of embedded massive compact cluster precursors and the observed gas-free clusters ($t_{\text{age}} = 1-2\Myr$), it can be assumed that the embedded phase is short, probably lasting $<1\Myr$.

Given the absence of observed timescales for the embedded phase, we modeled two cases: $t_{\text{emb}}=1\Myr$ (C1) and $t_{\text{emb}}=0\Myr$ (C0), see also Table~\ref{tab:set-up}, the latter being representative for a cluster with a very short embedded phase $t_{\text{emb}} \ll 1\Myr$. The consequences of these choices will be analysed in Sect.~\ref{sec:discussion}. Table 1 gives also the simulation parameters of two of the extended clusters modeled in  Vincke et al. 2016, for comparison. 
      
      The gas expulsion process brings the cluster out of equilibrium leading to members becoming unbound. 
However, as the SFE is quite high, this gas expulsion is only a secondary cause for cluster expansion for models C0 and C1. The main reason why the clusters expand by a factor of $10-20$ is the ejection of stars from the densest cluster regions \citep{Pfalzner_Kaczmarek_2013b} .

      % Table 3.
%\begin{table}[h!]

\begin{table}[b]
  \begin{center}
\caption{Compact cluster model set-up and dynamical timescales.}
\begin{tabular}[t]{l|lrrrrr|rrr} \tableline \tableline
%\tablewidth{0pt}
\hline 
	  model     & represents &$N_{\text{stars}}$ & $N_{\text{sim}}$ & SFE   & $r_{\text{hm}}$  & $t_{\text{emb}}$ & $M_{\text{stars}}$ & $M_{\text{cl}}$ & $t_{\text{dyn}}$ \\
\hline
		    &		&	   &                  &       & [pc]             & [Myr]            & [$\MSun$]          & [$\MSun$]       & [Myr]            \\ \hline
	  C0        &  Westerlund 1 & $32\,000$          & 10               & 0.7   & 0.2              & 0.0              & $18\,839$          & $26\,913$       & 0.01             \\
	  C1        &  Westerlund 1 & $32\,000 $          & 10               & 0.7   & 0.2              & 1.0              & $18\,824$          & $26\,891$       & 0.01             \\ \hline 
	E52        &  NGC 2244 & $32\,000$          & 9               & 0.3   & 1.3             & 2.0              & $18\,852$          & $62\,842$       & 0.1           \\ 
     E2         &  ONC      &    $4\,000$          &  94              &  0.3   &  1.3             &   2.0              &  $2\,358.1$         &   $7\, 860.3$     &   0.33  \\ 
\tableline 
\end{tabular}
	\vspace{0.5em}
	\tablecomments{Col.~1 gives the model name, $N_{\text{stars}}$ (Col.~2) is the number of stars in the model, $N_{\text{sim}}$ (Col.~3) the number of simulations in the simulation campaign, SFE (Col.~4) the star formation efficiency, $r_{\text{hm}}$ (Col.~5) the initial half-mass radius of the stellar and the gas component, $t_{\text{emb}}$ (Col.~6) the duration of the embedded phase, $M_{\text{stars}}$ (Col.~7) the stellar mass of the cluster, $M_{\text{cl}}$ (Col.~8) the total cluster mass (stars + gas), and $t_{\text{dyn}}$ (Col.~9) the dynamical timescale, see text for calculation.}
	\label{tab:set-up}
  \end{center}
      \end{table}
    
      During the simulations, we record for each fly-by the relevant parameters (time, duration, periastron, distance of primary to cluster center, mass ratio of encounter partners). This information is then used to post-process the data to obtain the effect of such a fly-by on the DPS.
For each model, simulations of only the first $3\Myr$ were performed.

It was assumed that each star was initially surrounded by a disK, meaning a 100\% initial disk frequency. This is not really observationally proven. The highest disk frequency observed for a compact cluster is approximately 30\% for NGC3603  \citep{Stolte_et_al_2015}. However, as disk destruction could be very rapid in such environments a 100\% initial disk frequency may still be a good assumption.  Even if the initial disk frequency should be considerable less, this would not be a problem for the results prsented here, as we are predominantly interested on the disk sizes.  Our results would then still hold for the stars that had initially disks.

    %- - - - - - - - - - - - - - - - - - - - - - - - - - - - - - - - - - - - - - - - - - - - - - - - - - - - - - - - - - - - - - - - - - - - - - - - - - - - - - - - - - - - - - - - 

    \subsection{Modelling the effect of a fly-by on the DPS size}
    \label{sec:modelling_the_effect_of_a_fly-by_on_the_disc_size}

We assume that initially each star was surrounded by a protoplanetary disk,  equivalent to a 100\% initial disk frequency. Observationally it is quite difficult to determine the initial disk frequency, especially in such dense environments. The highest disk frequency observed for a compact cluster is approximately 30\% for NGC 3603  \citep{Stolte_et_al_2015}. However, as disk destruction could be very rapid in such environments a 100\% initial disk frequency may be still a good assumption.  Even if the initial disk frequency would be considerable less, this would not be a problem for the results prsented here, as we are predominantly interested in the disk sizes.  In this case, our results would still hold for the stars that had initially disks. 

Similarily it is not straightforward what value one should assume for the primordial disk size.
Here we make one major assumption which currently is not testable by observations because there exist no observations of disk sizes in young  massive compact clusters.  This assumption is that disks around stars in compact clusters have initially the same properties concerning their size, mass, etc. as those
around nearby young stars that are part of a sparse cluster or associations. One has to keep in mind that those discs might differ in their properties because they normally form in an environment of higher gas and dust density and the embedded phase in compact clusters is much shorter. Consequences might be more massive compact disks, but there are equally strong arguments for less massive more extended disks. Without observations that is so far just speculation.  Therefore we assume a low-mass disk as characteristic for young stars in compact clusters.

In Taurus, a very sparse association, the disk size distribution peaks at $200\AU$, and disks of up to $700\AU$ are found \citep{Andrews_Williams_2007}. Observations show disk sizes of $27-500\AU$ in the ONC \citep{McCaughrean_Odell_1996, Vicente_Alves_2005, Bally_et_al_2015}, however, the ONC is denser than Taurus and at an age of at least $1\Myr$ it is questionable whether the measured disk sizes correspond to primordial ones. It is more likely that the disk size has already been altered due to photoevaporation or stellar fly-bys.
      
Disk size values found for example in Taurus should correspond to those unaffected by the environment. Therefore, we assume 200 AU as initial disk size for all stars in our simulations. We will see in the following that almost all disks in the compact clusters are stripped to sizes well below $100\AU$, so the result is completely independent from this initial choice. With this, we can separate processed from non-processed disks easily.

The question is whether and to which degree the DPS sizes can increase again after they have been reduced in size by a fly-by. There are various processes that could lead to DPS size growth, which in an extreme case could result in the disks becoming as large or even larger than they were initially. This is discussed in section 4.2.
      
    %- - - - - - - - - - - - - - - - - - - - - - - - - - - - - - - - - - - - - - - - - - - - - - - - - - - - - - - - - - - - - - - - - - - - - - - - - - - - - - - - - - - - - - - - 

    \subsubsection{DPS size determination after fly-bys}
    \label{sec:size_determination_after_encounters}

      For post-processing the fly-by data we make use of the  computational results from extensive numerical parameter studies, that determine DPS size as a function of the mass ratio $m_{12}=m_{2}/m_{1}$ and the periastron distance $r_{\text{peri}}$ \citep[][]{Breslau_et_al_2014} and the inclination \citep[][]{Bhandare_Breslau_Pfalzner_2016} during the fly-by. A detailed description of the different treatment of coplanar and inclined fly-bys is given in appendix A.
            
      In contrast to previous studies 
\citep[see e.g.][]{Clarke_Pringle_1993,Heller_1995,Hall_1997,Portegies_Zwart_2015,Vincke_Breslau_Pfalzner_2015}, Bhandare et al. (2016) implicitly took into account randomly orientated DPS,  which are a realistic situation for stellar clusters.
      They find that non-coplanar encounters have a still considerable effect on the DPS  size.
A database of the compuational results can be found at http://www3.mpifr-bonn.mpg.de/encounter-properties/.
      Averaged over all inclinations - including the coplanar prograde and coplanar retrograde case, they find the following approximate dependence:

      \begin{equation}
	r_{\text{disk}} = 1.6 \cdot r_{\text{peri}}^{0.72} \cdot m_{12}^{-0.2},
	\label{eq:bhandare_disc_size}
      \end{equation}

 where $r_{\text{disk}}$ is smaller or equal to the DPS  size before the encounter $r_{\text{previous}}$ (all sizes and distances are in AU). 

 Eq. 2 holds for mass ratios in the range 0.3 to 50, however, is unfortunately not straightforwardly applicable to our study. The reason is that their fit-formula focuses on very close to penetrating ($r_{\text{peri}} \leq r_{\text{previous}}$) encounters with the largest periastron distances included  5$\times r_{\text{init}}$.       
      In stellar clusters, even in the most compact ones, fly-bys with even larger periastron distances are the most common \citep[see also][]{Scally_Clarke_2001, Olczak_Pfalzner_Spurzem_2006}.      
      For this reason, we set up additional simulations analog to the ones by \cite{Bhandare_Breslau_Pfalzner_2016} with a focus on distant fly-bys. 
      We obtained a slightly different fit formula for the DPS size after an fly-by with ($m_{12}$, $r_{\text{peri}}$) averaged over all inclinations, that gives a better fit for distant fly-bys:
      
      \begin{equation}
	r_{\text{disc}} = \begin{cases} 
                               (1.6 \cdot m_{12}^{-0.2} - 1.26 \cdot m_{12}^{-0.182}) \cdot r_{\text{peri}}  & \mbox{for } r_{\text{disk}} < r_{\text{previous}} \\ 
				      r_{\text{previous}},                                                                                    & \mbox{for } r_{\text{disk}} \geq r_{\text{previous}}. 
	                         \end{cases}
	\label{eq:vincke_disc_size}
      \end{equation}
      
      A detailed description of the simulations, the data, and the fit are given in the appendix \ref{app:average_disc_size_after_randomly_orientated_fly-bys}. Equation~\ref{eq:vincke_disc_size} enables us to indirectly study randomly orientated fly-bys without having to simulate DPS and clusters simultaneously \citep[as done by][]{Rosotti_et_al_2014} which would not be possible for groups of $32\,000$ stars over timescales of $3\Myr$.
            
      % Figure 1.
      \begin{figure}[t!]
	\centering
	\includegraphics[width=0.45\textwidth]{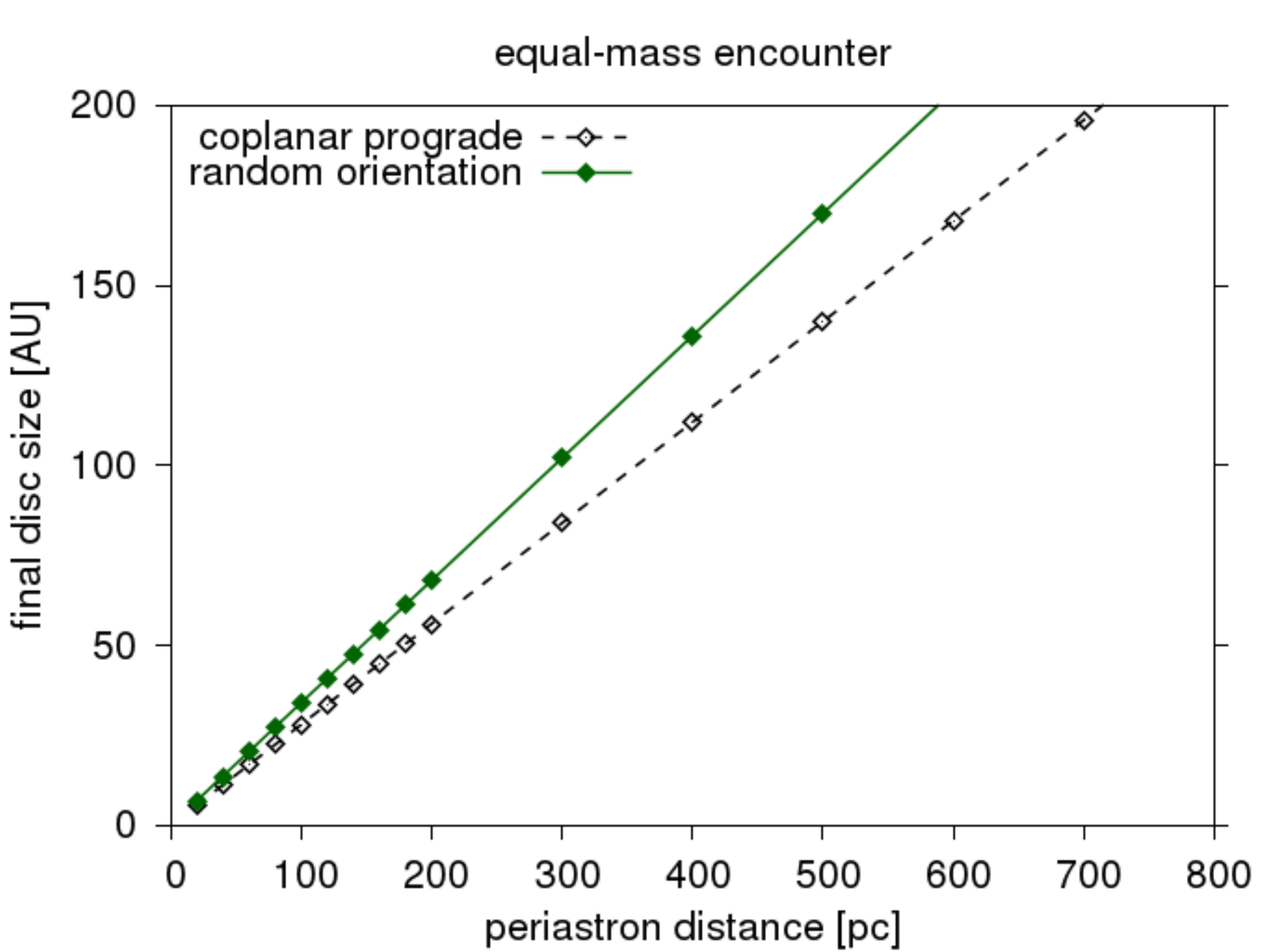}
	\caption{Final  DPS  size as a function of the periastron distance for coplanar, prograde fly-bys \citep[black open diamonds,][]{Breslau_et_al_2014} and randomly orientated fly-bys given by Eq.3 between two stars of equal mass ($m_{12}=1$).}
	\label{fig:influence_equal_mass}
      \end{figure}
      
      Previous simulations treated all fly-bys as coplanar and could only give upper limits for the effect of fly-bys, as such they overestimated it.
      However, Fig.~\ref{fig:influence_equal_mass} shows that there is a difference between coplanar and inclined fly-by, but it is relatively small. For example, a $200\AU$-sized disk would be truncated to $84\AU$ by a fly-by at a periastron distance of $300\AU$, whereas the disk size after a randomly orientated encounter would be $102\AU$, which is about $20\%$ larger.
      This is also reflected in the overall median DPS  size in each cluster type  (see appendix \ref{app:average_disc_size_after_randomly_orientated_fly-bys}).
      In the following, unless stated otherwise, we will present the outcome of our simulations assuming randomly orientated fly-bys.
      Here we only take into account events that lead to a DPS -size reduction of at least $5\%$ ($r_{\text{disk}}/r_{\text{previous}} \leq 0.95$).

Comparing theoretical and observational disk sizes poses many difficulties. On the one hand there are different definitions of disk sizes in the theoretical treatment as well as in observations. This issue is discussed in detail in  \citet[][]{Breslau_et_al_2014} and we use their definition of steepest gradient at the outside of the disk, due to its similarity to most observational methods. On the other hand, observational disk sizes depend on the wavelength range of the observations, whether one looks at the size of the gas or dust disk, the development stage of the disk and many other parameters  \citep{balog:16,marel:18}. Given the scarcity of disk size measurements in compact clusters, we just take the given data as their face value.  However, when more data will be available in the future, this will require finer specification.

We neglect effects other than fly-bys that could potentially lead to DPS size changes for example, viscous spreading or mass transport between disks, meaning we assume that the disk size remains constant throughout our simulations unless altered by a consecutive fly-by \citep[cf.][]{Rosotti_et_al_2014}.
 For a detailed discussion of potential DPS-size changing processes other than fly-bys see Sect.~\ref{sec:discussion}.

\section{Results}
\label{sec:results}
    
  The dynamical evolution of the compact clusters, that might develop into long-lived open clusters, differs considerably from that of the short-lived extended clusters/association that dominate the solar  neighborhood.
  It has been long expected that in such environments interactions are strong and very frequent and as such have a significant influence on the DPS. In the following we quantify the effect on the DPS sizes.

  %---------------------------------------------------------------------------------------------------------------------------------------------------------------------------------

  \subsection{Cluster evolution}
  \label{sec:cluster_evolution}

    % Figure 2.
    \begin{figure}[t!]
      \centering
      \includegraphics[width=0.45\textwidth]{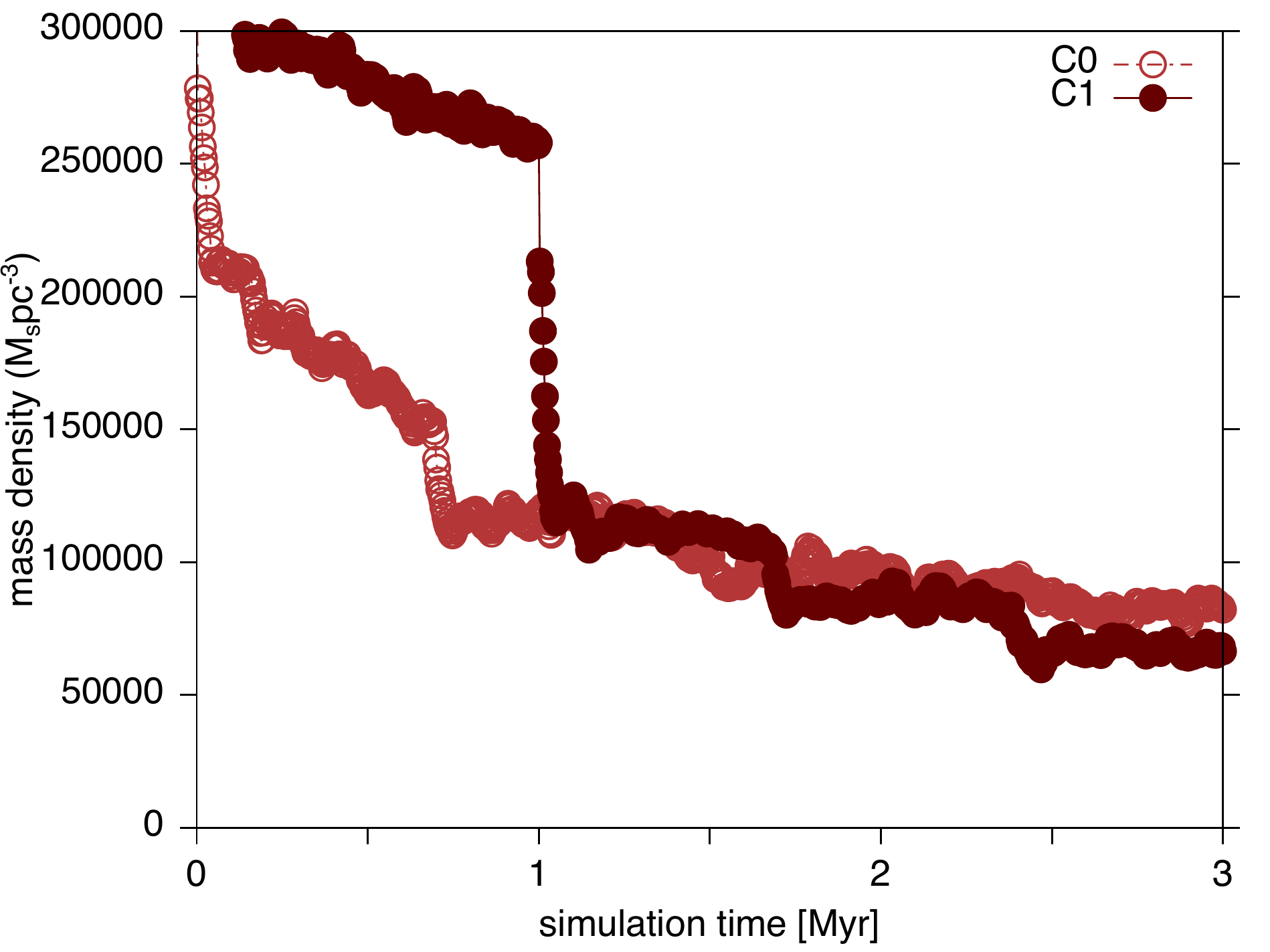}
      \caption{Mass density within the half-mass radius ($r_{\text{hm}}=0.2\pc$) as a function of time for the compact clusters (C0: light red circles, C1: dark red dots). The vertical line depicts the point in time of gas expulsion for \mbox{model C1} ($1\Myr$).}
      \label{fig:cluster_density}
    \end{figure}
    
    First, we want to look at the cluster evolution over the first $3\Myr$.  As mentioned in the introduction, in contrast to associations, compact clusters have a $70\%$ SFE, so the gas loss on its own only leads to a slight increase in the clusters size, but it is rather the ejection of stars in the dense cluster center that is responsible for the strong increase in cluster size \citep{Pfalzner_Kaczmarek_2013b}.
    In \mbox{Fig.~\ref{fig:cluster_density}} the stellar density development of the clusters within a sphere of their half-mass radius ($0.2\pc$) as a function of time is depicted for our compact models C0 and C1 (light red circles and dark red dots, respectively). Note that here the total system mass is taken into account, meaning the gas plus stellar mass. The gas expulsion at $t$=0 for C0 and $t$=1Myr for C1 results therefore in a drop in mass density by 30\%. In addition, does the loss of stars due to close fly-bys lead to a steady decrease in stellar density. As the stellar density decreases, stellar ejections become less common leading to a gradual slowing down of the expansion process.

  %---------------------------------------------------------------------------------------------------------------------------------------------------------------------------------
  
  \subsection{Cluster density}
  \label{sec:cluster_density}

    The cluster density determines the degree of influence of the environment on the protoplanetary disks surrounding its members. Naturally, the number of fly-bys potentially changing the disk size in the compact clusters is high, about 3-4 fly-bys per star during a period of $3\Myr$. This is not much more than for an extended cluster of the same mass, which has 1-2 such fly-bys per star despite having a lower initial density of roughly $150\,000\rhoM$ within $0.2\pc$.

    %Figure 3.
    \begin{figure}[t!]
      \centering
      \def\stackalignment{l}
      \includegraphics[width=0.45\textwidth]{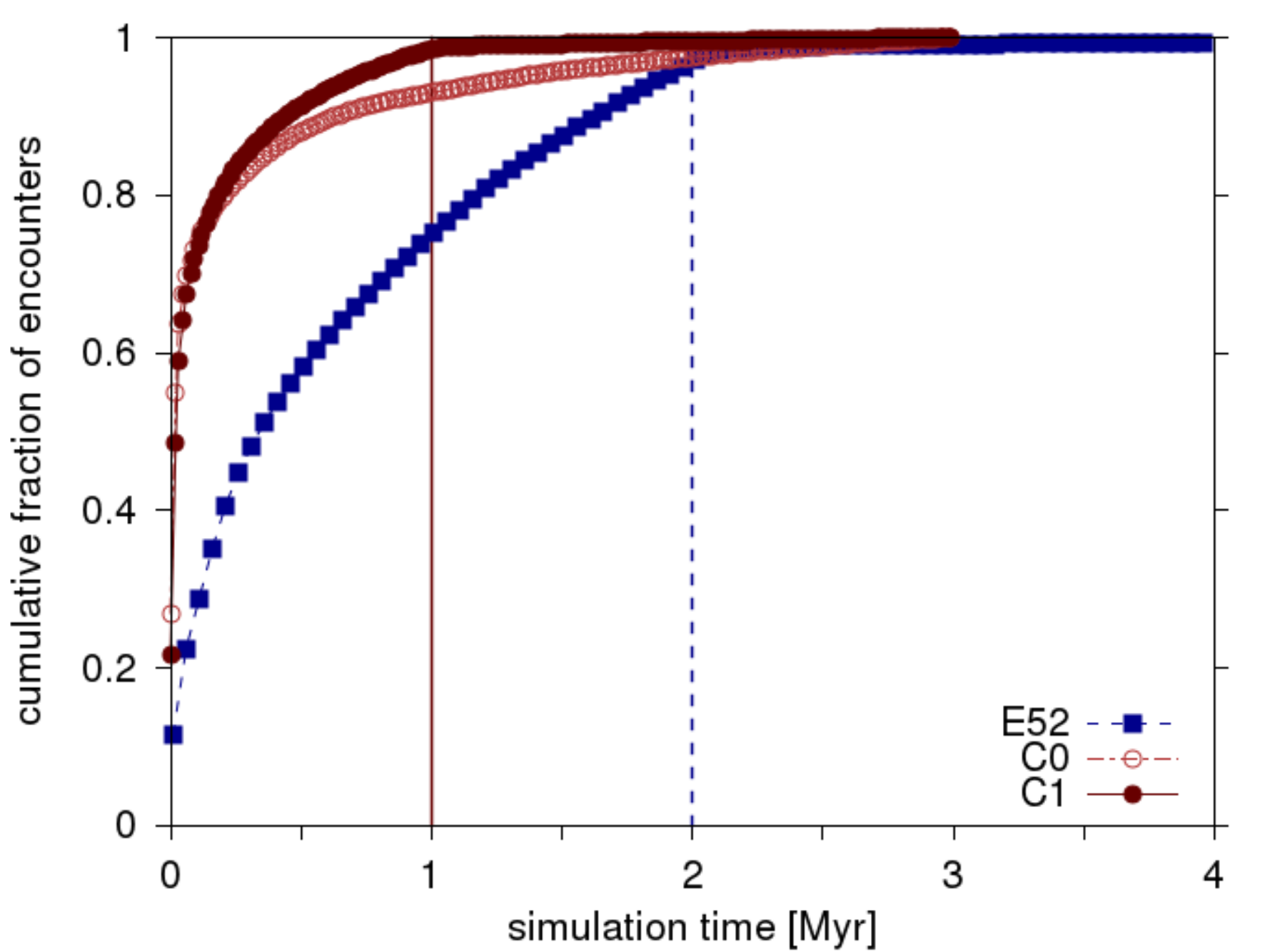}
      \caption{Cumulative fraction of randomly orientated encounters as a function of time for cluster models C0 (light red circles), C1 (dark red dots), E52 (dark blue, filled squares). The vertical black lines depict the point in time of gas expulsion for \mbox{model C1} ($1\Myr$, dashed) and \mbox{model E52} ($2\Myr$, solid).}
      \label{fig:dsc_enc_vs_time}
    \end{figure}
    
    The reason is the qualitative difference between these encounters: whereas in extended clusters the DPS size is reduced step by step, in compact clusters most disks experience very close fly-bys already within the first $0.1\Myr$, see \mbox{Fig.~\ref{fig:dsc_enc_vs_time}}.
    These very close fly-bys lead to such DPS disk sizes that the cross section for  a follow-up fly-by becomes small.
    As a consequence, the number of fly-bys that are actually necessary to produce such small DPS sizes is relatively small.

  %---------------------------------------------------------------------------------------------------------------------------------------------------------------------------------

  \subsection{Median DPS size}
  \label{sec:median_disc_size}
  
    % Figure 4.
    \begin{figure}[t!]
      \centering
      \def\stackalignment{l}
      %\topinset{\small \bfseries(a)}{\includegraphics[width=0.45\textwidth]{cluster_comparison__avg__mean_disc_size-200-par-incl-equ-mean_disc_size_edited.png}}{0.01cm}{0.01cm}
      \topinset{\small \bfseries(a)}{\includegraphics[width=0.45\textwidth]{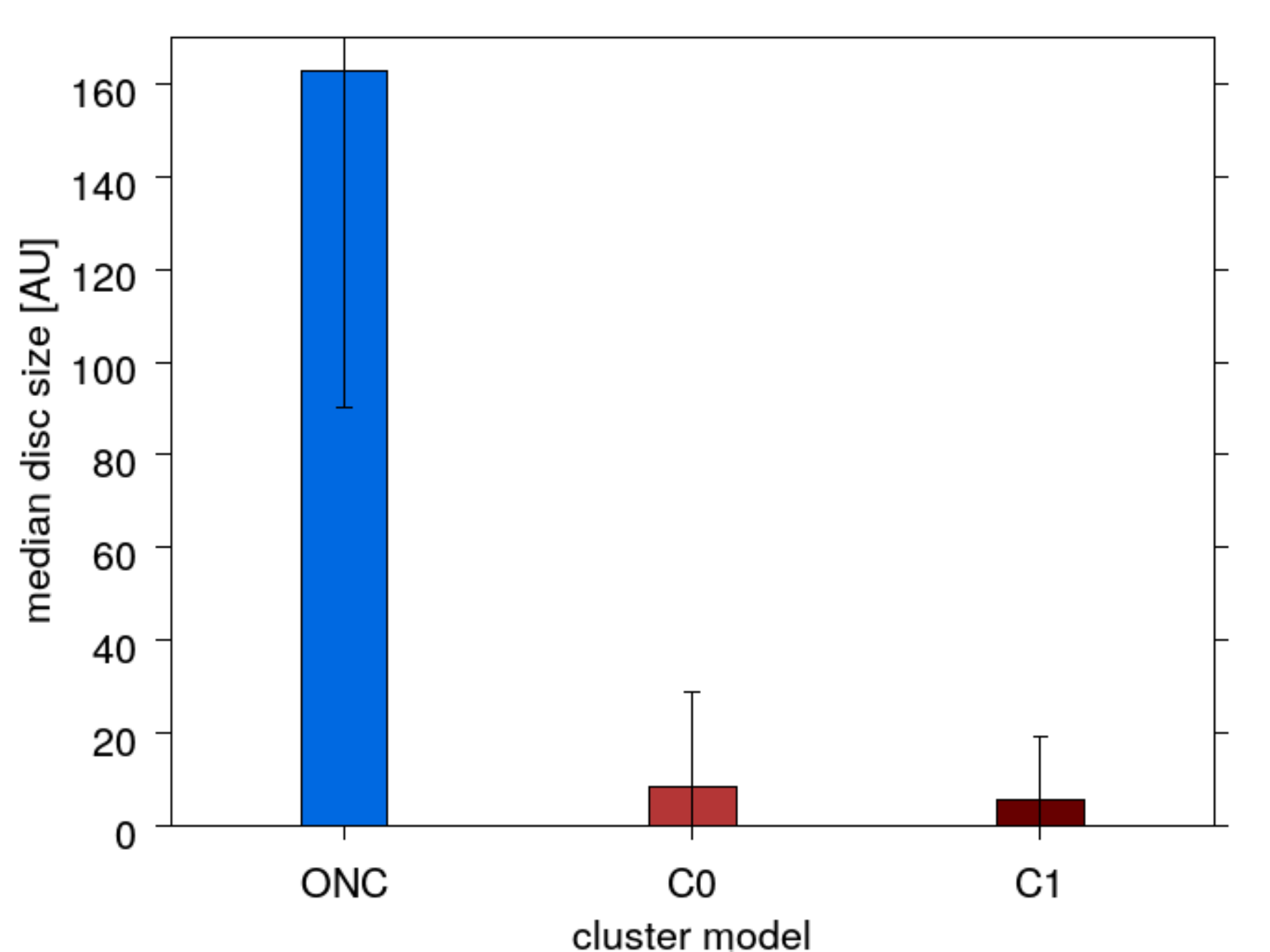}}{0.01cm}{0.01cm}
    %  \topinset{\small \bfseries(b)}{\includegraphics[width=0.45\textwidth]{cluster_comparison__avg__mean_disc_size-200-par-incl-equ-mean_disc_size_vs_distance_to_cluster_center_edited.png}}{0.01cm}{0.01cm}
\topinset{\small \bfseries(b)}{\includegraphics[width=0.45\textwidth]{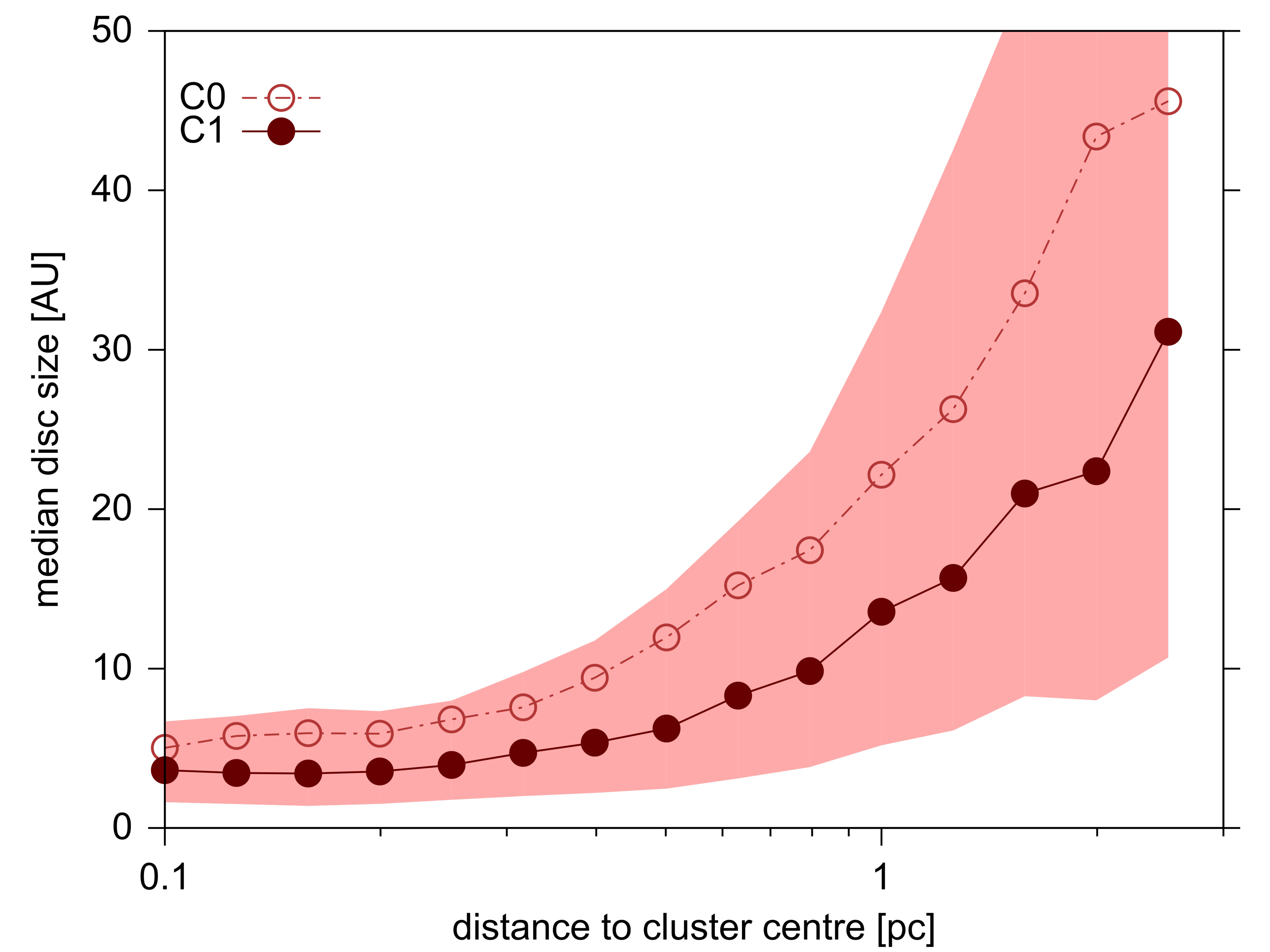}}{0.01cm}{0.01cm}
      \caption{a) Median DPS size at the end of the embedded phase forC0,  C1 $\&$ the ONC at the end of the simulations within the half-mass radius of the cluster with eroors indicated by the bars. b) Median DPS size as a function of the distance to the cluster center after $3\Myr$, Here the standard deviation is indicated by the shaded area.  Distance bins containing less than 10 stars for the whole simulation campaign were excluded.}
      \label{fig:median_disc_size_vs_model_and_distance}
    \end{figure}
    
    Next we study the DPS-size development in the different cluster environments. As expected, the high density in the compact clusters leads to small protoplanetary disks \citep[see e.g.][]{Bonnell_et_al_2001,Vincke_Breslau_Pfalzner_2015}. Most of the change occurs in the early phases of the cluster development. At 3 Myr the mean DPS size is  $21\AU$ for model C0 and  $11\AU$  in case of model C1. 
 For model C1 most of the DPS changes happen during the embedded phase, at the end of which the mean DPS size is already $12\AU$.
    
 As mentioned above, an embedded phase of such compact clusters is probably shorter than $1\Myr$, such that the real mean DPS size can be expected to be in the range $12\AU$ to $21\AU$ by $3\Myr$.  Fig. 5a) shows that the mean disk size within the half-mass radius is even smaller, approximately 10 AU and 8 AU for C0 and C1, respectively. The interaction dynamics can vary considerabley between different realisations \citep{Parker}, this is reflected in the relatively large error bars in Fig. 5a), which correspond to the standard deviation of the values.  

The DPS sizes are relatively small, as for example our own solar system with a Neptune semi-major axis of $\sim 30\AU$ could not have formed from such a small disk. On average, the DPS sizes in compact clusters are considerably smaller than extended clusters. For our model E52 a factor of four larger values are obtained and for the ONC (see Fig. 4a, blue), we obtain a typical value of 160 AU for the mean disk size within the half-mass radius, which would be characteristic for the dense clusters in the solar neighborhood.
    
    As the density contrast between the cluster center and its outskirts is very high, one expects that more encounters take place in the cluster core than in the outskirts. 
    This is reflected in the median DPS size as function of the distance to the cluster center shown in Fig.~\ref{fig:median_disc_size_vs_model_and_distance}b). 
    For models C0 and C1 (red) the median DPS size is smaller than $10\AU$ close to the  cluster center. For model C0 the median DPS size in the cluster core ($<0.3\pc$)  is $~8\AU$ and for  \mbox{Model C1} it is $~5\AU$ at $0.3\pc$. On the other hand, the DPS sizes are only larger than 10 AU at distances larger than 0.5 pc for C0 and 0.8 pc for model C1.  Again the scatter in the obtained values is relatively large, but the general trend is always the same with smaller values closer to the cluster center.  As will be discussed in section 4, it is an open question whether planets can form 
in such small disks at all.    In comparison, for model E52 the median disk size even in the most central area is at least 20 AU \citep{Vincke_Pfalzner_2016}. For model C0 even the stars at the outskirts of the cluster ($>$10 pc), which  mostly become unbound and leave the clusters still have a median disk size of $46\AU$. This means that any star which has been part of a compact cluster, even just for a short time, bears its marks by its small DPS size.

  %---------------------------------------------------------------------------------------------------------------------------------------------------------------------------------

  %Figure 5.
  \begin{figure}[t!]
    \centering
    \def\stackalignment{l}
  \includegraphics[width=0.45\textwidth]{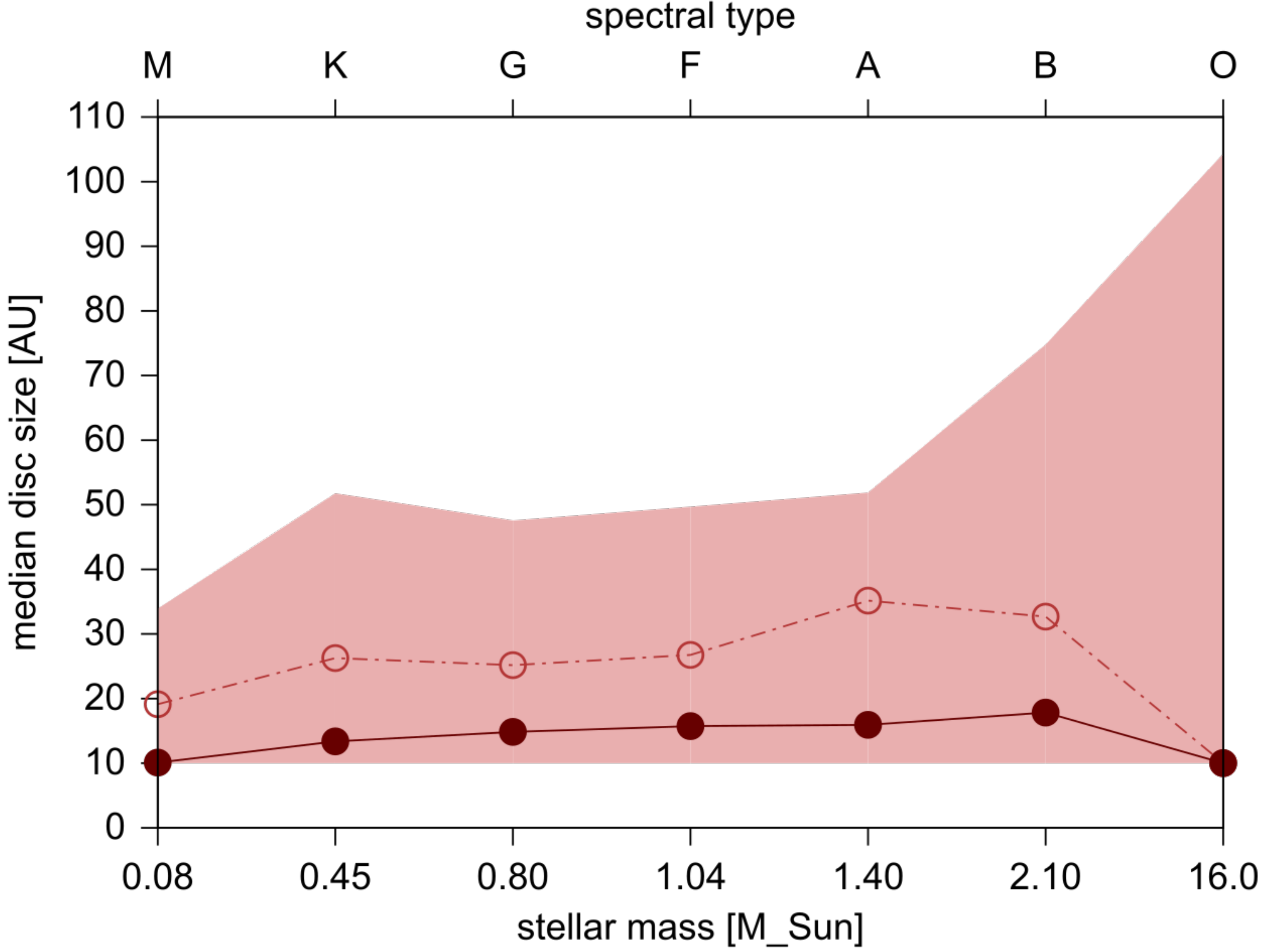}
    \caption{Median DPS size as function of stellar mass and stellar type for the three cluster models (C0: light red circles, C1: dark red dots, E51: dark blue squares).  The shaded region show the standard deviation of the values.}
    \label{fig:median_disc_size_vs_stellar_mass_and_type}
  \end{figure}

Next we discuss the question whether stars of different mass are affected to a different degree.
Fig.~\ref{fig:median_disc_size_vs_stellar_mass_and_type} shows that the final DPS size depends only slightly on the stellar mass.  From low mass-stars to B-stars there is only a slight increase in final mean DPS.
However, O-stars are much more affected by fly-bys than B-stars  as it is very conspicuous in model C0 where the average disk size of O-stars is $8\AU$, compared to $34\AU$ for the B-stars.
  The reason for the slight increase for M- to B-stars as well as the decrease for O-stars is the fly-by statistics. M-stars are most common and are therefore involved in most fly-bys, whereas the very massive ($\gg 20\MSun$) stars act as gravitational foci, therefore undergoing many strong fly-bys. A similar effect for the disk frequency has been noticed before, e.g. by \cite{Pfalzner_Olczak_2007}. However, for very large masses the statistics is less good as there are relatively few very high-mass stars.

 We assumed that initially all stars had the same disk size independent of the stellar mass, although theoretical \citep[see e.g.][]{Vorobyov_2011} and observational reasons support a slight dependence of the disk size on the stellar mass. It is often assumed that the disk size increases with stellar mass, ranging from roughly $100\AU$ for stars with $0.08\MSun$ up to nearly $1\,000\AU$ for solar-like stars, \citep[see e.g.][]{Vorobyov_2011}. Such an initial dependence would not alter the results presented here because all final mean DPS sizes are much smaller than 100 AU.

  %---------------------------------------------------------------------------------------------------------------------------------------------------------------------------------
  
  \subsection{Disk-size distributions in different environments}
  \label{sec:disc-size_distributions_in_different_environments}

    % Figure 6.
    \begin{figure}[t!]
      \centering
      \def\stackalignment{l}
  %\topinset{\small \bfseries(a)}{\includegraphics[width=0.45\textwidth]{cluster_comparison__avg__disc_size_change-200-par-incl-equ-disc_destroying_enc_vs_dist_to_cluster_center.png}}{0.01cm}{0.01cm}
  \topinset{\small \bfseries(a)}{\includegraphics[width=0.45\textwidth]{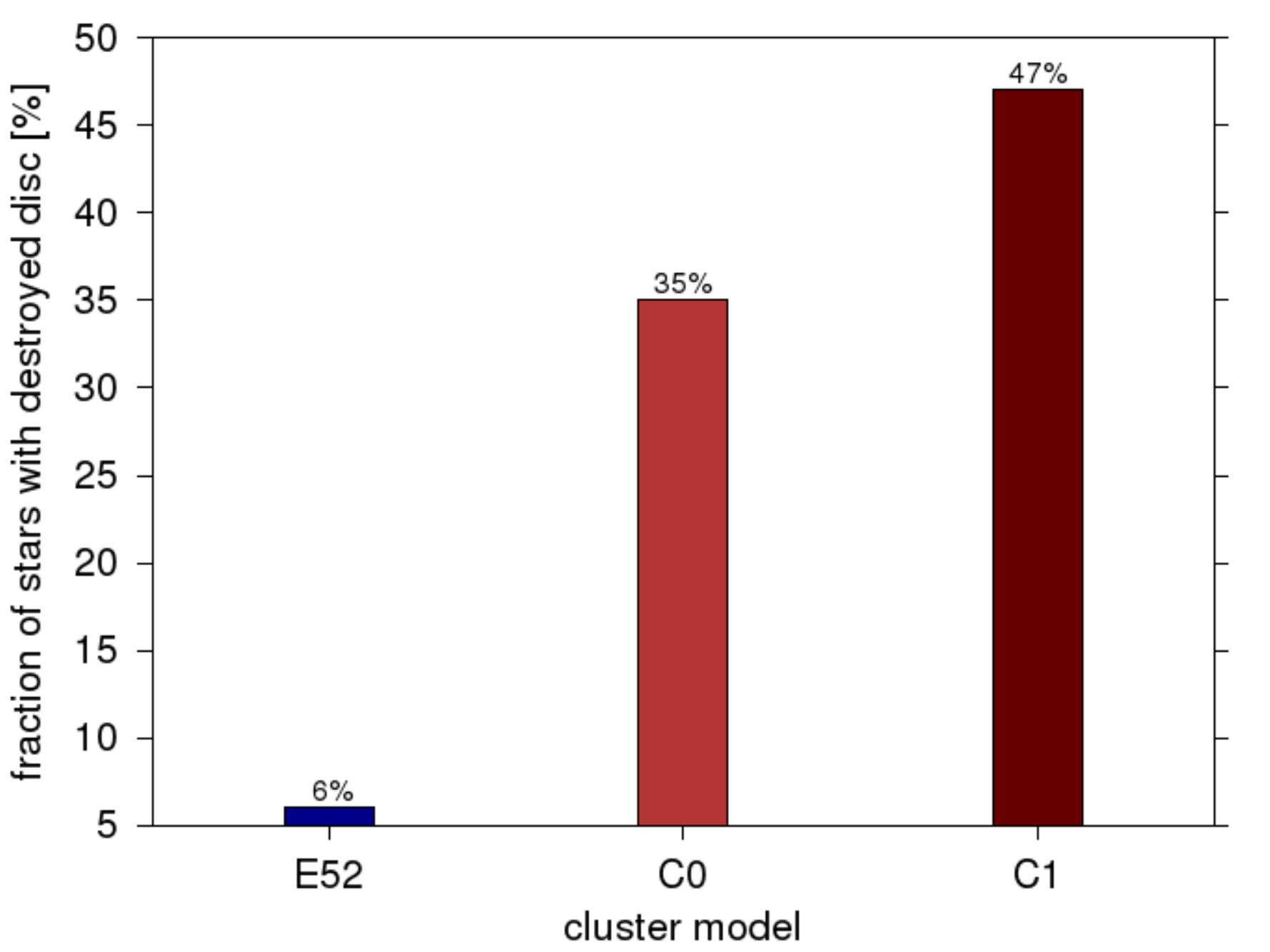}}{0.01cm}{0.01cm}
    %  \topinset{\small \bfseries(b)}{\includegraphics[width=0.45\textwidth]{cluster_comparison__avg__disc_size_change-200-par-incl-equ-disc_size_lt_100_AU.pdf}}{0.01cm}{0.01cm}
\topinset{\small \bfseries(b)}{\includegraphics[width=0.45\textwidth]{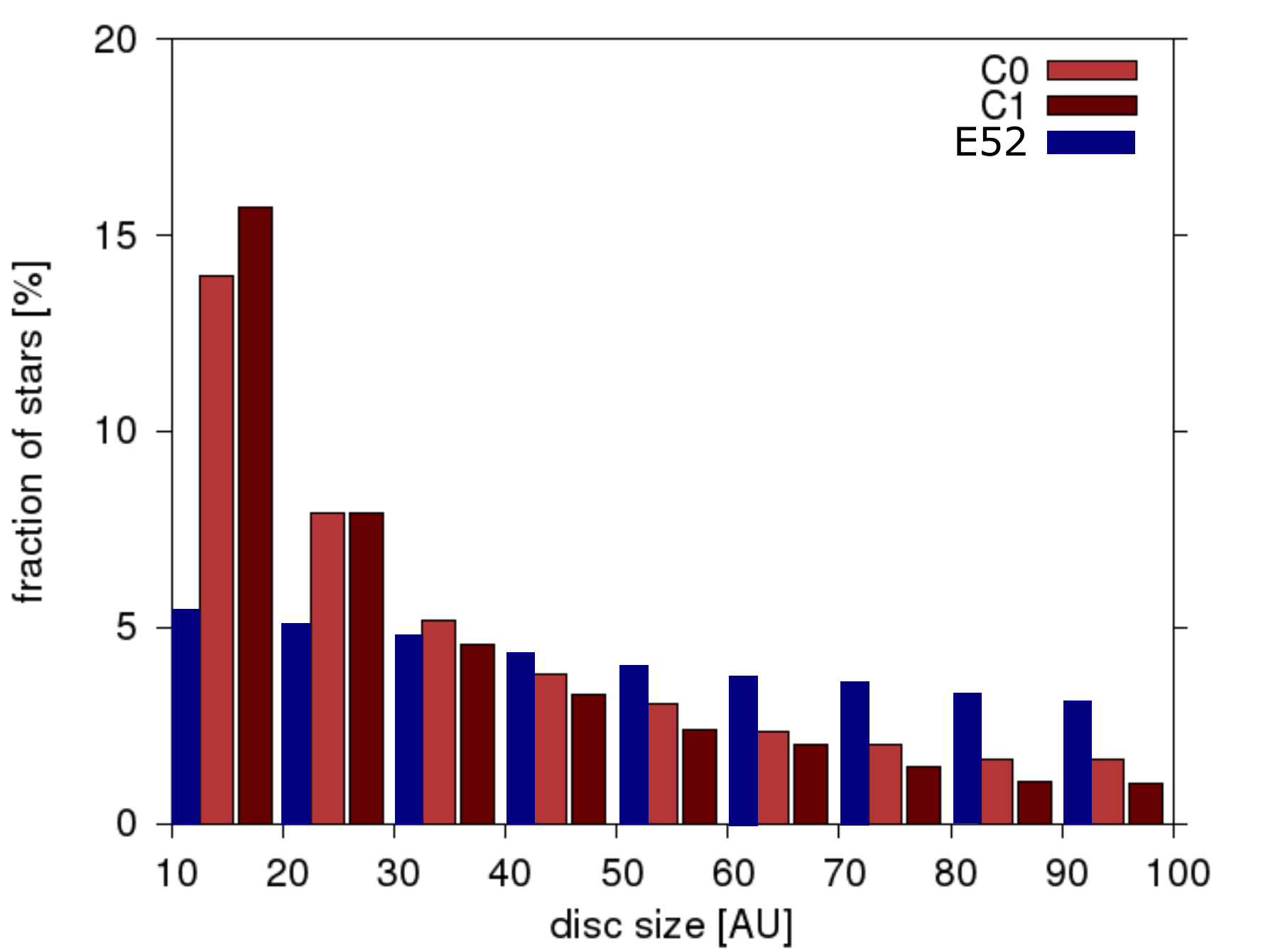}}{0.01cm}{0.01cm}
      \caption{a) Fraction of stars with disk sizes smaller than $10\AU$ at end of simulation in the whole cluster. b) Disk-size distribution within an artificial field of view of $3.0\pc$ at the end of the simulations ($t_\text{sim}$) for the different cluster models for disk sizes smaller than $100\AU$. The compact clusters are depicted in red (C0: light red, C1: dark red), the extended cluster (E52) in dark blue.}
      \label{fig:disc-size_distribution}
    \end{figure}

    How does the disk size distribution look like in such compact clusters? Figure~\ref{fig:disc-size_distribution} shows the size distributions for the two compact cluster models. Here we try to mimic observations, and consider only stars within $3\pc$ - the typical field of view (FOV) in observation at 200 to 400 pc distance,  for example, with the Spitzer telescope. Outside this radius, member determination is very difficult due to the usually high stellar back-/foreground densities. 

Surprisingly, disk sizes smaller than $10\AU$ are the most common ($35\%$ and $47\%$, respectively) in such compact clusters, see Fig.~\ref{fig:disc-size_distribution}a). If the disks are cut down to such a small size before planet formation took place, it is unlikely that there is enough material left for gas giants to form by accretion afterwards, because the remaining mass in the disk is relatively small. However, if one of the formation channels is hierarchical fragmentation, gas giants might still form in such hostile environments. In some cases there might be sufficient material to form terrestrial planets by accretion, but this requires further investigation. 

Large disks are quite rare, only as much as $23\%$, or even $ 14\%$ for model C1, of disks are larger than $100\AU$ and they are mostly located at the outskirts of the clusters. In the extended cluster only about $37\%$ of all stars are smaller than $100\AU$ after $10\Myr$. Even though the simulation time was much longer, the majority of disks remains large and only very few are destroyed ($6\%$).

  %Figure 7.
%  \begin{figure}[t!]
  %  \centering
   % \def\stackalignment{l}
   % \includegraphics[width=0.45\textwidth]{cluster_comparison__avg__disc_size_change-200-par-incl-equ-disc_destroying_enc_vs_dist_to_cluster_center.png}
   % \caption{Fraction of stars with disk sizes smaller than $10\AU$) at end of simulation in the whole cluster. The compact clusters are depicted in red (C0: light red, C1: dark red), the extended cluster (E52) in dark blue.}
    % \label{fig:destroyed_discs}
  %\end{figure}

\section{Discussion}
\label{sec:discussion}

  In our  cluster simulations we have made a number of assumptions and simplifications, namely we assumed (1) instantaneous gas expulsion, (2) did not include mass segregation and (3) did not include primordial binaries. We will discuss the potential influence of each of these in the following section. 

  %---------------------------------------------------------------------------------------------------------------------------------------------------------------------------------
  
  \subsection{Cluster dynamical evolution}
  \subsubsection{Cluster dynamics}
  \label{sec:discussion_gas_expulsion}

    We investigated the effect of the duration of the embedded phase by comparing models $t_{\text{emb}} = 0$ (C0) and $t_{\text{emb}} = 1\Myr$ (C1).
    Model C0 can rather be used to set constraints to disk-size distributions of clusters which are embedded for less than $1\Myr$ (C1).
    The duration of the embedded phase does make a difference for the resulting disks and possibly forming planetary systems: the fraction of small disks ($\le 20\AU$) is much larger in the embedded model and the median disk sizes differ by up to a factor of two, see Figs.~\ref{fig:disc-size_distribution} and \ref{fig:median_disc_size_vs_model_and_distance}b.
    
    We assumed the gas expulsion itself to happen instantaneously. This is, at least for the investigated very massive and dense clusters, a justified approximation. In reality, the gas expulsion is thought to last a few dynamical timescales which are of the order of $0.01\Myr$ for the compact clusters and $0.1\Myr$ for the extended model E52, so modelling the gas expulsion instantaneously is reasonable. 
    If the gas expulsion would last longer, the clusters would have enough time to adjust to the gas-mass loss, so less stars would become unbound. Additionally, the mass density would stay higher for a longer timespan leading to a higher encounter frequency and thus smaller disks. 
  
  %---------------------------------------------------------------------------------------------------------------------------------------------------------------------------------
  
   \subsubsection{Mass segregation}
   \label{sec:discussion_mass_segregation}
    
 In our study, the masses and positions of stars in the clusters were picked out randomly from the respective distributions disregarding initial mass segregation. Although a lot of clusters seem to be mass segregated, it is still under debate whether this is an initial property or a consequence of dynamical evolution. If the number density would remain the same but with the more massive stars concentrated in the center, the encounter rate in the cluster core would increase because of the larger gravitational focusing by the massive stars. Hence, the disks around the most massive stars would be smaller or could even be destroyed.
    
  %---------------------------------------------------------------------------------------------------------------------------------------------------------------------------------
   
  \subsubsection{Binaries}
  \label{sec:discussion_primordial_binaries}
  
    In our simulations all stars were initially single and binaries were not included in the set-up. The reason why we did neglect binaries, is that the data for the effect on the disk of binaries is only available for single cases, but no systematic parameter studies like the ones required here do exist. However, observations show that a large portion of stars are binaries \citep{Koehler_et_al_2006, Duchene_Kraus_2013}.
    This means that the fraction of stars surrounded by disks presented here is an upper limit, as more disks would be destroyed when - or not even form - as part of a binary. The periodic interactions between a disk and its binary star would have to be investigated in more detail to make further predictions about its size and structure.

    The inclusion of binaries would also affect the cluster evolution: \cite{Kaczmarek_PhD_2012} demonstrated that binaries lead to an accelerated cluster expansion on timescales of a few Myr, but as most fly-bys that lead to disks-size truncation occur during the first Myr this should not alter the results presented here about the disk sizes.
    In addition, observations and simulations show that the most massive stars are most probably part of binary systems. This increases gravitational focusing which in turn might result in even smaller disk sizes.

In our simulations some captures of stars into binaries do occur, however, these are relatively rare processes. They are resolved in the cluster simulation. However, in the effect on the disk size they are treated like that of a fly-by at the resulting binary separation. Like in the above case of primordial binaries, this leads to underestimating the effect and would require further investigation in the future.

  \subsubsection{Parabolic encounters}

  \label{sec:discussion_discs}
    
    All fly-bys were assumed to be parabolic, that is the eccentricities are $e=1$. The encounter eccentricity depends on the cluster density \citep[][]{Olczak_Pfalzner_Eckart_2010, Olczak_et_al_2012}, especially in the case of very dense clusters investigated here where the real eccentricities can be fairly high due to the dominance of  three-body interactions. The topic of the dependence of the disk size on the eccentricity of the fly-by has been scarcely investigated in the literature. However, studies of the disk-mass loss indicate that hyperbolic fly-bys might be less destructive \citep{Pfalzner_2004}. Recent investagtion by  \citep{winter:18a} indicate that this is also true for disk sizes.

  \subsubsection{Stellar evolution}

We did not include stellar evolution in our simulations as we think that the influence on the results is limited.
The minimum mass for a star to become a supernova during the 3 Myr covered here is \mbox{70 \MSun}. Thus some of the
realisations could in principle experience a supernova explosion towards the end of the simulation. However, then
most of the disk truncation processes have already taken place, so that the influence of the supernova on the
cluster dynamics should be very small. To have a considerable influence on the cluster dynamics it would be required that the supernova explosion
would take place within the first Myr, but the required stellar mass would be about 400 \MSun.

Even if no supernovae explosion would take place, the mass loss of the massive stars even during their first few Myr might be important \citep{vink:15}.
We did not include mass loss in our simulations, because, first, the number of high-mass stars is very-low in comparison to the low-mass stars, 
thus their effect on the average disk size is low. Second, in clusters like the ONC the massive stars play a fundamental role in the dynamics as they act as gravitational focii. In this type of clusters mass loss of the massive stars might possibly play some role. However, for the much denser clusters, which we discuss here, the massive stars lose this role, as interactions with the lower mass members become much more common \citep[][]{Olczak_Pfalzner_Eckart_2010}. Basically there needs to be sufficient space around the massive stars to become a focus, if that space is filled up, the massive stars loose their special role as focii. Therefore, the resulting average disk size is dominate by far by the interactions between the low-mass stars. Thus even if the massive stars would lose considerable amounts of their mass, this would not influence the result in any sizeable way. Third, if we take 10$^{-7}$\MSun\ yr$^{-1}$ as an example, this would correspond to a 0.3 \MSun\ loss. A 100\MSun\ star would be reduced to a 99.7 \MSun\ star after 3 Myr, that makes hardly any difference on the size after a fly-by. Even the 3 \MSun difference expected with a 10$^{-6}$\MSun\ yr$^{-1}$ mass loss would change the disk size value by less than 2AU even in a very close encounter, which is negligible in an averaging computation over 32000 stars.  A constant mass loss rate of 10$^{-5}$\MSun\ yr$^{-1}$ would be a problem, especially for stars of 30 \MSun\ and smaller, as they would have dispersed completely by then. However, we doubt that high-mass stars continuously have such high mass loss rates over the entire 3 Myr modeled here. It is much more likely that the mass-loss rate varies strongly with time, even showing burst like is known for accretion. Nevertheless the role of mass loss in such massive clusters should be considered in future studies.

  %---------------------------------------------------------------------------------------------------------------------------------------------------------------------------------

  \subsection{Other influence on DPS size}

The results from the effect of fly-bys described in section 2.2.1.  can be used under certain conditions to describe the effect on the disk and also on planetary systems. The difference is that for disks the simulation particles are representative for the mass distribution of the dust, whereas for planetary systems they represent the parameter space where planets potentially move on circular orbits before the fly-by.

The question is whether and to which degree the DPS sizes can change after they have been reduced in size by a fly-by. There are various processes that could lead to DPS size growth, which in an extreme case could result in the disks becoming as large or even larger than they were initially. We have to distinguish between the processes that play a role in the disk phase and those that are only relevant in the phase where a planetary system has already formed.

\subsubsection{Disk phase}

The main process that can change the disk size in addition to fly-bys is external photo-evaporation \citep{Haworth:16}. The radiation from the massive stars is strong enough to ionize the material in the disk and gradually material is removed form outskirts of the disk. Therefore, the disk size can be reduced by external photo-evaporation.  Thus, the above results can be considered as upper limits of the disk size, because  external photo-evaporation could lead to a further decrease in disk size.  Unfortunately, the degree and especially the timescale on which external photo-evaporation happens is less constrained than the gravitational effect of fly-bys \citep{Gorti:16}. In addition, comparison between the expected effect of external photo-evaporation and observations indicate that the real effect is smaller than that predicted by theory \citep{Gorti:16}.  Therefore, the average effect of external photo-evaporation in clusters is  currently unknown.  It is also important to note, that external photo-evaporation is only efficient when the cluster is no longer heavily embedded in gas. This means that one would only expect additional disk reduction by photo-evaporation after the embedded phase has ended.

However, there are not only processes that could lead to further disk size reduction, but also the ones that potentially would lead to a size increase.  One of them is viscous spreading during the disk phase, which could lead to an increase in disk size before and after the fly-by. The disk size growth before the fly-by does not influence our results because basically all disks are affected by fly-bys  to such a degree that they become smaller than 100 AU independently of their pre-fly-by size. Potentially more important is the viscous spreading that might take place after the fly-by.  However, considerable disk size increase can only happen if (i) the disk is relatively massive, so that the disk has a sufficiently high viscosity and (ii) the gas stays in disks for a sufficiently long time. As we have no observational information about the disks masses in compact clusters, we have no idea whether they are more or less massive than around nearby stars. Assuming it is the same as around nearby young stars, this corresponds to a typical disk mass $m_d \sim 0.01 M_s$. For such low-mass disks considerable viscous spreading ($>$20 AU) is noticeable after $\approx$ 5-10 Myr. However, in compact clusters the average lifetime of disks is very short, at most 1-2 Myr, so that disk spreading is $\ll$ 20 AU. This means that the potentially resulting planetary systems are mostly  $\ll$ 40 AU. 

  \subsubsection{Planetary-system phase}
 There are basically two processes that could lead to changes in the system size of an already formed planetary system after the fly-by - capture of one or more planets or excitation of planets onto eccentric orbits by planet-planet scattering. 

Generally capturing requires a close fly-by, but in this case not only material is captured but at the same time the disk size is reduced.  The material captured from another star tends to go onto highly eccentric orbits with the periastron being relatively small. The latter means that in most cases $r_{peri}(captured) < r_d $.  As soon as there are objects on eccentric orbits, it is no longer straightforward how to define the disk size. If we consider the periastron distance, then only captured matter does not influence the disk size. However, if one considers the outermost periastron as relevant, then an increase in DPS size could in principle happen. 

How common would be such a capture event? Capturing a planet from another star requires a relatively close fly-by and many fly-bys that lead to DPS truncation are too distant to lead to capturing. In other words, capturing is less common than DPS disk size decrease.  This again implies that planet capturing processes are most common during the early cluster phases and are even rarer than DPS decrease in later phases.  Therefore, the question is whether planets that could be potentially captured have already formed during the first $\approx$5 Myr, so that they could lead to a larger DPS size than expected from our results. It is still an open question what is the shortest time span required for planet formation, with estimates ranging from 1 to 10 Myr. The additional difficulty is that the context here implies captured planets that would have been originally orbiting at relatively large distances ($> 50 AU$) from their previous host star,  because only this lightly bound matter can be swapped between stars during a fly-by. According to the standard accretion model, planet formation proceeds slower in the outer disk regions. Therefore the only alternative would be that these planets formed early on due to gravitational instability in a relatively massive disk. In summary, there might be some cases where planet capture can lead to a larger DPS afterwards, but these cases are likely quite rare meaning that they should not significantly alter the results presented above.

The only process which could lead to considerably larger DPS in compact clusters than the ones discussed above, is in our opinion planet-planet scattering. Here long-term interactions between the planets orbiting a star can lead to the ejection of planets or the excitation of the orbit of one or more members of the system to a larger distance from the host star. How common such a process is depends on the compactness of the original planetary system. Again it is not known how compact planetary systems around stars and particularly around cluster stars typically are. Even if we assume that compact systems are common and excitation to more distant orbits happen often, this would probably not reflect in a larger observed DPS size in the near future. The reason is that mostly the least massive planet is excited onto a wider generally eccentric orbit.  Unfortunately 
in the foreseeable future it will not be possible to detect low-mass planets moving on wide orbits in mostly fairly distant compact clusters.   

In summary, the  above presented results should not only be representative for the situation at an age of 10 Myr but also for the long-term appearance of planetary systems in such compact clusters. 

 \subsection{Comparison with other models}

As mentioned in the introduction most simulations so far concentrated on DPS sizes in less dense clusters. These are more typical in the solar neighborhood and as such more observational data exist for comparison. 
\citet{winter:18b} investigate clusters spanning from low-density clusters to those similar to the ones studied here. They did not take into account the cluster development, meaning the cluster density stays constant for the entire 3 Myr they model in their Monte Carlo approach. However, they treat the effect of external photo-evaporation in addition. 
Taking both processes into account, \citet{winter:18b} obtain a mean disk size of approximately 85 AU at a cluster age of 3 Myr for their low-density model (model D): The density in this model  is slightly lower than the one by \citet{Vincke_Pfalzner_2016}. However, both these values for typical disk sizes in low-density clusters are significantly larger than the value of 21 AU, we obtain for the compact clusters considered here. This means there is a significant difference between the typical DPS sizes  in the typical solar neighborhood clusters and the ones likely to develop into long-lived open clusters. 

\citet{Pfalzner:18a} modeled one specific open cluster, namely, M 44. This cluster has probably had a similar size as the one modeled here, but an approximately 10 times lower mass to start with and therefore an about 10 times lower stellar density.  Therefore it might be considered a prototype for many of the open clusters with lower masses. As the stellar density is lower than in the case modeled here, one can expect that the disk sizes should on average be slightly larger. Or, that the relative number of very small disks should be slightly smaller.  Fig. 6 in \citet{Pfalzner:18a} confirms this expectation. There one can see that 13\% of all disks in M 44 are smaller than 10 AU compared to about 35\% - 47 \% in the case studied here. 

\citet{winter:18b}  considered also compact clusters similar to the ones we consider here. They find that generally external photo-evaporation dominates as disk truncation process over the effect of fly-bys. Does this mean fly-bys are only a second order effect for the disk size? Not really, they specifically point out that their model only applies to the situation when the cluster is no longer strongly embedded, as only then external photo-evaporation can act to its full extent.  We saw in section 3 that 80\% of the disk truncating fly-bys happen during the first 0.2 Myr of cluster development, meaning during the deeply embedded phase i.e. the phase when external photo-evaporation does not act yet. The mean disk size at the end of the embedded phase is well below 30 AU. We have to compare that to the value provided by \citet{winter:18b}, which give a mean disk size of  35-38 AU for their highest density cluster (model D). This means that the disk sizes at the end of the embedded phase are already smaller than what one would expect to happen due to photo-evaporation in the consecutive 3 Myr. In other words, fly-bys dominate the disk truncation process during the embedded phase. It is not clear what happens afterwards, as the results of \citet{winter:18b} can not simply be transferred to these smaller disks. A future study is required that includes not only the effect of fly-bys and external photo-evaporation, but also the embedded phase and the cluster expansion process. 

The clusters investigated here differ from extended clusters not only through their much higher density but also by having  a star formation efficiency that is at least 60\%. In a recent study, \citet{wijnen:17} compare the effect of dynamical encounters with that of a special model of  face-on accretion and ram pressure stripping. They find that the latter are dominating as long as the total cluster mass in stars is $\leq$ 30\% regardless of the cluster mass and radius. In other words, encounters dominate over face-on accretion and ram pressure in the compact clusters which we have investigated here. However, in the early star formation this effect could be important and should be investigated in a dedicated study.

\section{Summary and conclusion}
\label{sec:summary_and_conclusion}

  For a long time it was unclear whether planets could form and survive in the dense stellar environment of open clusters. However, during the past few years several protoplanetary disks  and about 20 planets have been found in open clusters showing that planets can indeed withstand such harsh environments.
 Some open clusters are probably older counterparts of young compact clusters like Arches or Westerlund 2. Here, we have studied the influence of stellar fly-bys on the size of protoplanetary disks  and eventually forming planetary systems by performing simulations of such young, compact clusters. We have considered two models, one with a very short (C0), the other one with  a $1\Myr$-long embedded phase (C1).
  Starting with an initial disk size of $200\AU$, we find that in both cases stellar fly-bys are responsible for significant changes in the disk size. After $3\Myr$ all disks  are reduced in size by fly-bys, with $77\%$ and $86\%$ being smaller than $100\AU$ in C0 and C1, respectively.
  However, what is most interesting is that disk sizes $<10\AU$ are the most common in such environments, $35\%$ and $47\%$ of all disks  are smaller than $10\AU$ in model clusters C0 and C1, respectively. This corresponds to mean disk sizes of $21\AU$ for C0 and $11\AU$ for C1.
    
Disks  which are reduced to sizes smaller than $10\AU$ have a relatively small mass and their structure is very complex \citep[][]{Breslau_et_al_2014}. It is not certain that the remaining material would be identified as a disk any more by observations.  As their structure differs considerably from a flat Keplerian disk, it is not clear what effect that feature has on the planet formation process.   It could either prevent it all together, because at least in the outer areas the matter is unevenly distributed, or it could actually accelerate dust growth due to an induced increase in collision frequency. This has to be considered in more detail in future work. 

 Here we only consider the effect of fly-bys on the disk size, however in such dense and massive clusters the strong radiation from the massive stars - external photoevaporation - can lead to additional disk size reduction. However, in contrast to fly-bys, external photo-evaporation works most efficiently at the end of the embedded phase; the relative importance of these two effects should be determined in the future.
 
  % Table 1.
  \begin{table}[!b]
  %  \centering
    \caption{Observed properties of disks  in open clusters.}
    \begin{tabular}[b]{llllll} \toprule 
		      & \multicolumn{2}{l}{Cluster properties}  & \multicolumn{2}{l}{disk properties}  & Ref.      \\ 
      Name            & $t_{\text{cl}}$  & $M_\text{cl}$        & $\#$  & $r_{\text{disc}}$            &           \\
		      & [Myr]            & [$10^3 M_{\odot}$]   &       & [AU]                         &           \\ 
      %---------------------------------------------------------------------------------------------------------------------------
      NGC~2362        & $4-5$            & $>0.5$               & 4     & $6.2-40.9^{a)}$              & (1), (2)  \\
      h/$\chi$Persei  & $13 \pm 1$       & $\gtrsim 4/3$        & 10    & $1.0-38.4$                   & (3), (4)  \\ %\bottomrule
      %---------------------------------------------------------------------------------------------------------------------------
    \end{tabular}
    \vspace{0.5em}
    \tablecomments{Column~1 indicates the cluster name, $t_{\text{cl}}$ its age and $M_\text{cl}$ its mass, the number of observed disks  is shown in Col.~4 and the (dust) disk radius $r_{\text{disc}}$ in Col.~5, Col.~6 gives the references.
    References: (1) \cite{Dahm_2005}, (2) \cite{Currie_et_al_2009} and ref. therein, (3) \cite{Slesnick_Hillenbrand_Massey_2002}, (4) \cite{Currie_et_al_2008}.
    Comments: a) Lower dust radius for \mbox{200~K} and upper dust radius for \mbox{120~K}, respectively.}
    \label{tab:observed_disks _in_open_clusters}
  \end{table}

The observational statistics of DPS sizes in compact and/or open clusters are still scarce so comparing our results with observations has to be done with great care. 
The estimated sizes vary in the range of a few AU up to roughly $40\AU$, see Table~\ref{tab:observed_disks _in_open_clusters}. We can conclude that the overall median DPS sizes found in our simulations ($11\AU$ and $21\AU$) agree surprisingly well with the sizes of DPS found in open clusters ($1-40\AU$, see Tables~\ref{tab:observed_planetary_systems_in_open_clusters} and \ref{tab:observed_disks _in_open_clusters}. )

  It should be emphasized that the clusters in Table 4  differ considerably from those in clusters in the solar   neighborhood, such as the ONC. The Arches cluster at 7~kpc might be a bit far away, but in clusters like NGC 3606 the disk sizes might be possible to determine with ALMA. However, above results do not only restrict the disk sizes in such compact clusters, but also the sizes of the planetary systems that can form and survive in such hostile environments. Our simulations predict also that planetary systems in such environments will have usually sizes of $<10\AU$ and it will be basically impossible that systems with sizes of $100\AU$ will form and survive in open clusters. 

 Observations show disk sizes of $27-500\AU$ in the ONC and $\ge 3-12\AU$ the Arches cluster
 \citep{Eisner_et_al_2008, Stolte_et_al_2010}. A comparison with the solar system puts these sizes in perspective. Neptune orbits at an approximate distance of $30\AU$ from the Sun. However, every fourth protoplanetary disk and/or planetary system in such open cluster progenitors has a size of less than $30\AU$ (model C1). A size of $21\AU$ corresponds roughly to the distance to Uranus and $11\AU$ is just outside Saturn's orbit.
  
Another question is how many of these small disks  contain sufficient material to form planets or even planetary systems. As the 23 observed planets in open clusters show, obviously at least some do. Two of those planets even form a planetary system consisting of a Hot Jupiter and a very eccentric Jupiter-like planet.
However, we expect that these planetary systems differ considerably from those found around field stars. So systems with planets on orbits $100\AU$ wide should be very rare or even non-existent unless they were captured from another star \citep{jilkova:15}. 

In addition, many of the systems should have sharp outer edges like our own solar system  \citep{Pfalzner:18b} and the orbits of the outer planets should be often very eccentric and inclined relative to the inner system.
  Actually one planetary system found so far in an open cluster - Pr0211 in M44 - looks just like we would predict from our simulations: the outer planet moves on a highly eccentric orbit with $a_{\text{pl}} \leq 5.5\AU$ \citep{Pfalzner:18a}. This is exactly the kind of planetary system which we would expect to dominate in such open cluster environments. \\

\acknowledgements 
 The authors would like to thank Asmita Bhandare for providing and extending the star-disk encounter simulations. Our additional thanks go to the anonymous reviewer, for a thorough and insightful review that has led to a substantial improvement of the manuscript.

\appendix
  
  \section{Inclined vs. coplanar, prograde fly-bys}
  \label{app:inclined_vs_coplanar_prograde_fly-bys}

  % Figure A.1.
  \begin{figure}[b!]
    \centering
    \def\stackalignment{l}
    \topinset{\small \bfseries(a)}{\includegraphics[width=0.45\textwidth]{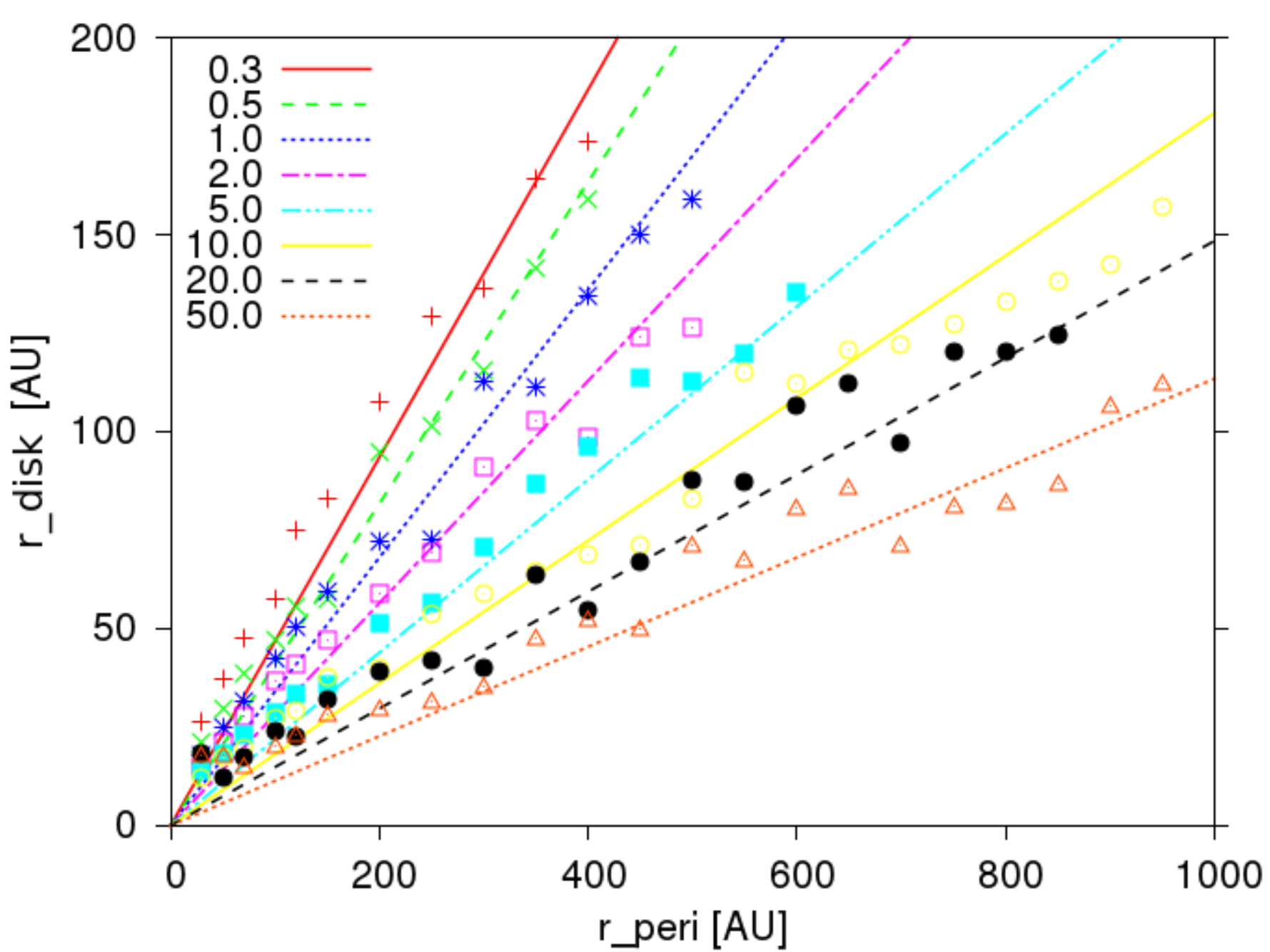}}{0.01cm}{0.01cm}
    \topinset{\small \bfseries(b)}{\includegraphics[width=0.45\textwidth]{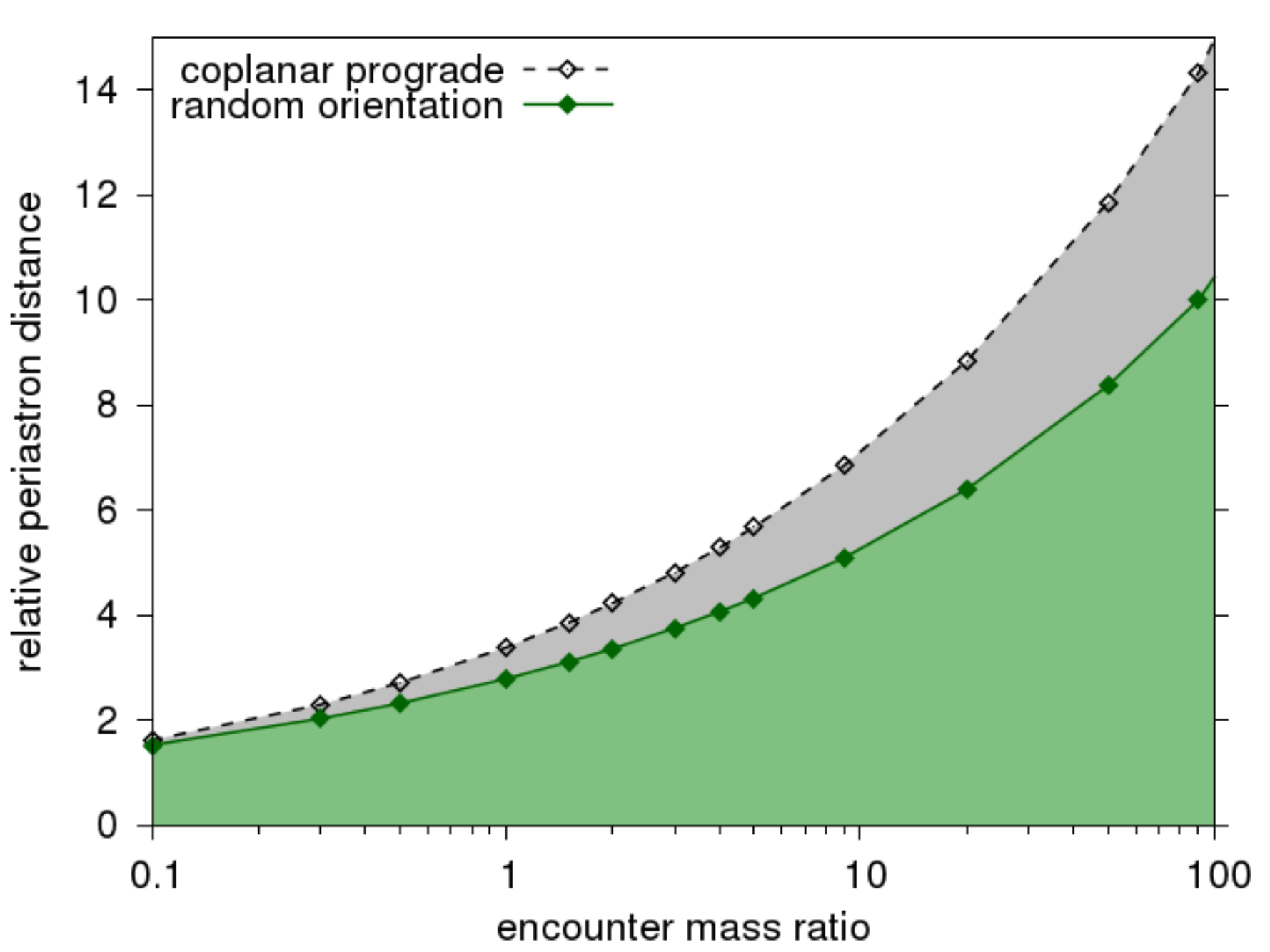}}{0.01cm}{0.01cm}
    \caption{a) Combined data sets of \cite{Bhandare_Breslau_Pfalzner_2016} and new simulations including best fit to data given by Eq.~\ref{app_eq:vincke_disc_size}. The different colours depict different encounter mass ratios from $m_{12}=m_2/m_1=0.3$ (red) to $m_{12}=50$ (orange). For details about simulations and excluded data points, see text. 
    b) Parameter pairs (encounter mass ratio and relative periastron distance $r_{\text{peri}}^{\text{rel}} = r_{\text{peri}}/r_{\text{previous}}$) leading to $5\%$ reduction in disc-size for coplanar, prograde encounters (black open diamonds, Eq.~\ref{app_eq:breslau_disc_size}, \cite{Breslau_et_al_2014}) randomly orientated encounters used in this work (green filled diamonds, Eq.~\ref{app_eq:vincke_disc_size}). The light grey and green areas depict the parameter space in which the disc size is reduced for the respective fly-by types.}
    \label{app_fig:new_simulations_and_influence}
  \end{figure}
  
  The size of a disk around a star with mass $m_1$ after a fly-by with a star of mass $m_2$ at a periastron distance of $r_{\text{peri}}$ has been recently investigated numerically in two large parameter studies. 
  The first by \cite{Breslau_et_al_2014} {\bf  analyzed}  a thin disk of mass-less tracer particles around a star after it was perturbed by a second star on a coplanar, prograde orbit with respect to the disk.
  They found a simple description of the disk size after such a fly-by depending on the mass ratio of the two stars ($m_{12} = m_2/m_1$) and the periastron distance:
  
  \begin{equation}
    r_{\text{disk}}^{\text{copl}} = \begin{cases} 0.28 \cdot r_{\text{peri}} \cdot m_{12}^{-032}, & \mbox{if } r_{\text{disk}} < r_{\text{previous}} \\ 
				    r_{\text{previous}},                                          & \mbox{if } r_{\text{disk}} \geq r_{\text{previous}}, 
		      \end{cases}
    \label{app_eq:breslau_disc_size}
  \end{equation}
  
  where $r_{\text{previous}}$ is the disk size previous to the fly-by in AU.
  
  This parameter study was extended by \cite{Bhandare_Breslau_Pfalzner_2016} by including fly-bys of different inclinations, that is different angles between the disk plane and the plane of the perturber's orbit. Averaging over all inclinations for one set of ($m_{12}$, $r_{\text{peri}}$), they obtained the following fit formula:
  
  \begin{equation}
    r_{\text{disk}}^{\text{avg}} = \begin{cases} 1.6 \cdot r_{\text{peri}}^{0.72} \cdot m_{12}^{-0.2}, & \mbox{if } r_{\text{disk}} < r_{\text{previous}} \\ 
				    r_{\text{previous}},                                               & \mbox{if } r_{\text{disk}} \geq r_{\text{previous}}, 
		      \end{cases}
    \label{app_eq:bhandare_disc_size}
  \end{equation}
  
  This formula describes fly-bys with mass ratios of 0.3 to 50.0 and up to periastron distances of five times the initial disk size $r_{\text{init}}$. However, the focus of the fit was disk-penetrating  and close fly-bys (roughly up to two times $r_{init}$).
  In stellar clusters, even in the most compact ones, fly-bys with larger periastron distances and/or high mass ratios are still the most frequent type of fly-by. Despite the distance, these fly-bys can lead to disk truncation if the mass ratio is large \citep[see also][]{Scally_Clarke_2001, Olczak_Pfalzner_Spurzem_2006}.
  To be able to describe such distant fly-bys more precisely, we have performed additional encounter simulations, analog to the ones by \cite{Bhandare_Breslau_Pfalzner_2016}: a disk of $10^6$ mass-less tracer particles was set up around a star and a second star passed by, removing particles from the disk and reshaping the remnant disk.
  The disk size after such an encounter was chosen to be the steepest point in the time-averaged density distribution. The initial disk size was set to $200\AU$, mass ratios of 0.3 to 50 and all inclinations ($0^{\circ}-180^{\circ}$) were covered. Periastron distances between $400\AU$ and $1\,000\AU$ were covered in steps of $50\AU$ to obtain a better resolution than before in this parameter range that is important for our simulations, see Fig.~\ref{app_fig:new_simulations_and_influence}.  In this case the points for closer fly-bys where excluded from the fit.

  \begin{equation}
    r_{\text{disk}} = \begin{cases} (1.6 \cdot m_{12}^{-0.2} - 1.26 \cdot m_{12}^{-0.182})\cdot r_{\text{peri}}, & \mbox{if } r_{\text{disk}} < r_{\text{previous}} \\ 
				    r_{\text{previous}},                                                         & \mbox{if } r_{\text{disk}} \geq r_{\text{previous}}. 
		      \end{cases}
    \label{app_eq:vincke_disc_size}
  \end{equation}
  
  Note that self-gravity as well as viscosity within the disk were neglected. More details about the simulation set-up, the disk-size determination and the influence of the above made assumptions are given in \cite{Breslau_et_al_2014} and \cite{Bhandare_Breslau_Pfalzner_2016}.
  
  %-----------------------------------------------------------------------------------------------------------------------------------------------------------------------------------

  \section{Average disk size after randomly orientated fly-bys}
  \label{app:average_disc_size_after_randomly_orientated_fly-bys}

\begin{figure}
    \centering
\includegraphics[width=0.45\textwidth]{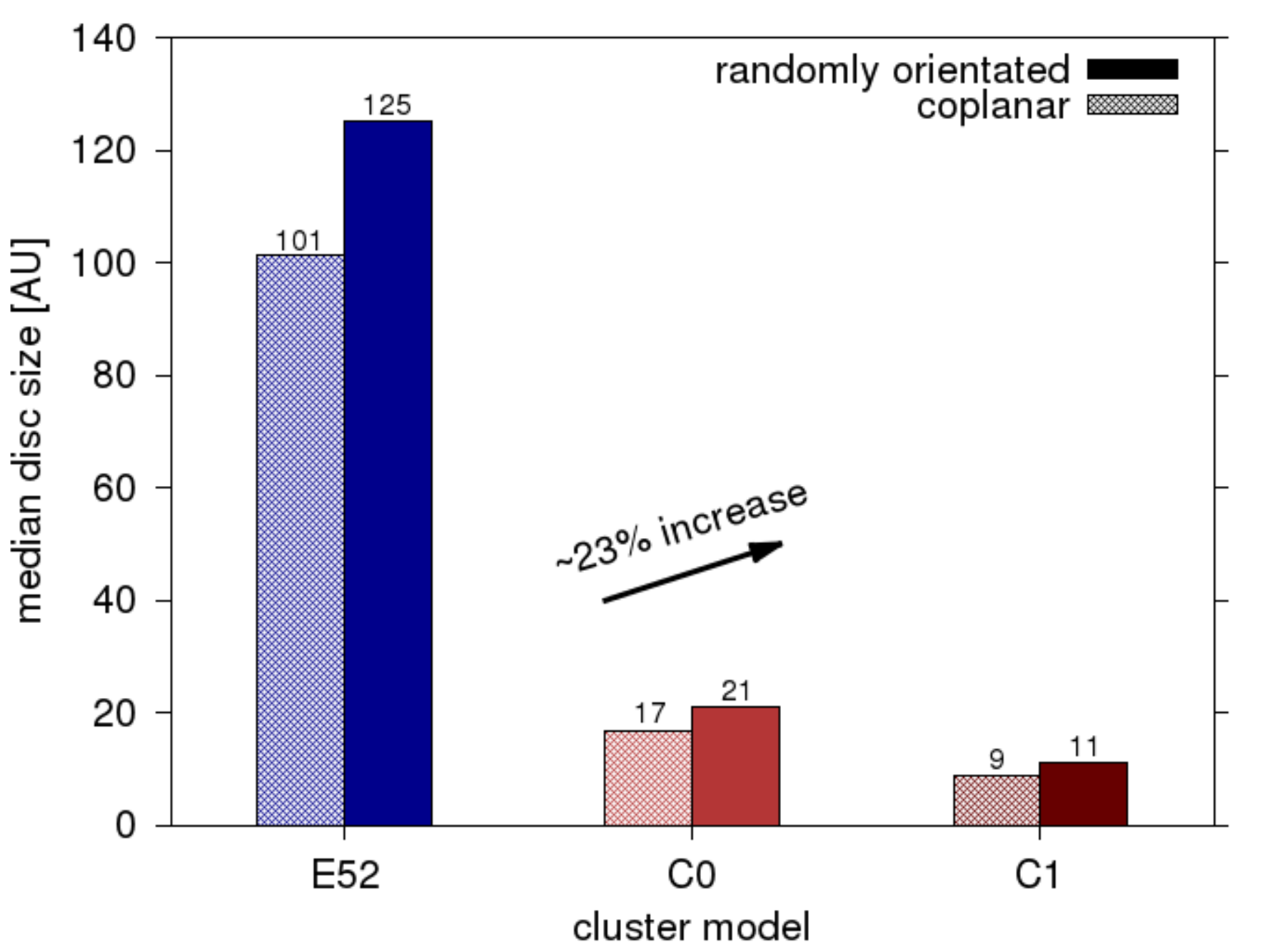}
    \caption{Median disc size after $10\Myr$ assuming all fly-bys to be coplanar and prograde (crossed) versus including all inclinations (filled) for the different cluster models.}
    \label{app_fig:coplanar_vs_inclined}
  \end{figure}

  With the fit formula above, we can now quantify the difference in disk size obtained assuming fly-bys to be coplanar to those where it is assumed that all inclinations are equally likely.
  Fig~\ref{app_fig:coplanar_vs_inclined} depicts the median disk size after $10\Myr$ where 1) all fly-bys were assumed to be prograde, coplanar, using the disk-size description by \cite{Breslau_et_al_2014} (crossed boxes), and 2) taking into account all inclinations using the average disk size as defined by Eq.~\ref{app_eq:vincke_disc_size} above (filled boxes).

  % Figure B.1.

  As expected the median disc size of randomly orientated fly-bys is larger than for purely coplanar, prograde ones in all cluster models. 
  The absolute difference is actually quite small, especially in the compact clusters. Nevertheless, in relative terms, the median disc size increases by $23\%$ comparing coplanar to randomly orientated fly-bys, independently of the cluster.
  
  %The discs are not only larger when the fly-bys are considered to be randomly orientated, but there are also more fly-bys taking place, especially in the extended cluster model E52 (about $18\%$ increase as compared to purely coplanar, prograde fly-bys). 
  %This is due to our definition of an encounter: the disc size should be altered by at least $5\%$, so as the discs stay larger after randomly orientated fly-bys, more follow-up fly-bys are capable of reducing the disc size even more. 
  %This effect is of minor importance in the starburst clusters, as the discs around $35\%$ (C0) and $47\%$ (C1) of all stars are destroyed and follow-up fly-bys are not registered in our statistics, see Fig.~\ref{fig:disc-size_distribution}.

%----------------------------------------------------------------------

% Bibliography.
% -------------
%\bibpunct{(}{)}{;}{a}{}{,} % to follow the A&A style

% bibliographystyle not needed for aa style because set by the aa documentclass
%\bibliographystyle{aa}

%\bibliography{local}

\end{document}